\documentclass[a4paper,fleqn,usenatbib]{mnras}
\usepackage{txfonts}
\usepackage[T1]{fontenc}
\usepackage{ae,aecompl}

\usepackage{graphicx}	% Including figure files
\usepackage{xcolor,ulem}
\usepackage{cancel}
\usepackage{slashed}
\usepackage{lipsum}
\usepackage{float}
\usepackage{relsize}
\usepackage{multirow}

\def\be{\begin{equation}}
\def\ee{\end{equation}}
\def\bea{\begin{eqnarray}}
\def\eea{\end{eqnarray}}
\def\bn{\begin{enumerate}}
\def\en{\end{enumerate}}
\def\bi{\begin{itemize}}
\def\ei{\end{itemize}}

\def\Eiso{E_{iso}}
\def\tv{\theta_v}
\def\tj{\theta_j}
\def\tp{t_{\rm peak}}
\def\ts{t_{\rm start}}
\def\tvj{\theta_{vj}}
\def\eB{\epsilon_B}
\def\eE{\epsilon_E}
\def\p{p}
\def\Msun{M_\odot}

\def\dg{^\circ}

\makeatletter

\newcommand{\Rmnum}[1]{\expandafter\@slowromancap\romannumeral #1@}
\makeatother

\title[Exploring afterglow parameter space]{Exploring Short-GRB afterglow parameter space for observations in coincidence with  gravitational waves}
\author[M. Saleem et al.]{
M. Saleem$^{1}$\thanks{E-mail: saleemc87@iisertvm.ac.in},
L. Resmi$^{2}$,
Kuntal Misra$^{3}$,
Archana Pai$^{4}$ and 
K. G. Arun$^{5,6}$
\\
$^{1}$Indian Institute of Science Education and Research Thiruvananthapuram, CET Campus, Trivandrum 659016 \\
$^{2}$Indian Institute of Space Science and Technology, Trivandrum. \\
$^{3}$Aryabhatta Research Institute of Observational Sciences, Nainital. \\
$^{4}$Department of Physics, Indian Institute of Technology Bombay, Powai, Mumbai 400076 \\
$^{5}$Chennai Mathematical Institute, Siruseri, 603103 Tamilnadu.\\
$^{6}$Institute for Gravitation and the Cosmos, Pennsylvania State University, State College, PA 16802.}

\date{Accepted XXX. Received YYY; in original form ZZZ}

\pubyear{2017}

\begin{document}
\label{firstpage}
\pagerange{\pageref{firstpage}--\pageref{lastpage}}
\maketitle

\begin{abstract}
{Short duration Gamma Ray Bursts(SGRB) and their afterglows are among the most promising electro-magnetic (EM) counterparts of Neutron Star (NS) mergers.} The afterglow emission is broadband, visible across the entire electro-magnetic window from $\gamma$-ray to radio frequencies. The flux evolution in these frequencies is sensitive to the multi-dimensional afterglow physical parameter space. {Observations of gravitational wave (GW) from BNS mergers in spatial and temporal coincidence with  SGRB  and associated afterglows can provide valuable constraints on afterglow physics.} 
We run simulations of GW-detected BNS events and assuming all of them are
associated with a GRB jet which also produces an afterglow, 
investigate how detections or non-detections in X-ray, optical and
radio frequencies can be influenced by the parameter space. 
We narrow-down the regions of afterglow parameter space for a uniform top-hat
jet model which would result in different
detection scenarios. We list inferences which can be drawn on the
physics of GRB afterglows from multi-messenger astronomy with
coincident GW-EM observations.
\end{abstract}

%<<<<<<< HEAD

%%%%%%%%%%%%%%%%%%%%%%%%%%%%%%%%%%%%%%%%%%%%%%%%%%

%%%%%%%%%%%%%%%%% BODY OF PAPER %%%%%%%%%%%%%%%%%%

\section{Introduction}
%(PARA 1: SHORT GRBS, PARA 2:INSTRUMENTS AND DETECTABILITY OF PROMPT AS WELL
%AS AGS, PARA 3: GW OBSERVATIOSNS AND INSTRUMENTS, PARA 4: JOINT SCENARIO, PARA 5:
%WHAT IS OUR AIM WHICH SHOULD INCLUDE BRIEF SUMMARY OF AG THEORY AND PARAMETERS, PAGE 6:
%PAPER ORGANIZATION)
%          \tc{ADD INSTRUMENTS, AFTERGLOW DISCUSSION, STUDY OF AFTEGLOW AND LITERATURE}  

%	\tc{[Resmi/Kuntal/Arun to correct] GW DETECTORS CURRENT STATUS}.
%        Coincident GW-SGRB
%        detections have been addressed in literature in various
%        contexts[\cite{siellez2013simultaneous},
%          \cite{regimbau2015revisiting},
%          \cite{clark2015prospects}]. There are previous studies which
%        address the joint detectability of GW and SGRB/afterglows. For
%        example, \cite{metzger2012most} address the E-n space of
%        EM-counterparts of the GW-detectable NS mergers. Recently,
%        \cite{lazzati2016off} explored the detectability of various EM
%        components including AG with fixed GRB parameters but
%        different viewing angles.  \cite{patricelli2016prospects}
%        investigate the detectability of prompt and afterglow with
%        Fermi in coincidence with BNS sources for different jet
%        opening angles. \tcb{Hotokezaka et al 2016} does
%        something. \cite{fong2015decade} does something and
%        \cite{feng2014detectability} also does something with ag
%        parameter space

{The most favoured progenitor model for short duration Gamma Ray Bursts (SGRBs) are mergers of two neutron stars (NS) or a black hole and a neutron star. The recent observation of gravitational waves from BNS merger GW170817\citep{GW170817} in spatial and temporal coincidence with GRB170817A has confirmed Binary Neutron Star (BNS) mergers to be one of the progenitors of SGRBs \citep{MMApaper,GRB+GW-2017}. As advanced Gravitational Wave (GW) detectors are coming online, in future more GW triggers will be followed up by EM instruments in various bands.} {Joint GW-EM detections will help us put constraints on the GRB parameters and improve our knowledge of the underlying physics of the source. \citep{arun2014synergy, bartos2013gravitational, GRB+GW-2017}}. %\textbf{add more citations from the bns event, like Kim et al, Granot et al, Hallinan et al, Troja et al etc. two are added}

{The origin of the long-duration GRBs has been firmly established to be the massive star collapse}  through their association with
stripped envelope Type-Ib/c supernovae \citep{Hjorth:2003jt}. {Until the recent joint detections of GW170817 and GRB170817A, there were no direct evidence for the progenitors of SGRBs.} The absence of an associated supernova in SGRBs was the most supportive evidence for the binary merger proposal \citep{Levan:2007zr, Fong:2016irn}. 
There were further additional observations consistent with the binary merger hypothesis. For
example, {the diversity in SGRB host galaxies, including both early and
late type galaxies, were indicative of the progenitors belonging to old
stellar populations \citep{Fong:2013eqa}}. The GRB positions showed a
statistically larger off-set from the photo-center of their host
galaxies, naturally explained from the natal kicks of Neutron Stars
\citep{Fong:2009bd}.  {In 2013,} the observation of a nearby short GRB 130603B showed an excess emission in near-infrared, consistent with the
expectations of a kilonova emerging from the r-process nucleosynthesis
of neutron rich material ejected from the merger \citep{Li:1998bw, Berger:2013wna, Tanvir:2013pia}.

Direct and indirect evidences associate GRBs to relativistic outflows \citep{Goodman:1986az, Paczynski:1986px, 1993ApJM, Frail:1997qf, Taylor:2004wd} collimated to narrow opening angles of a few degrees \citep{Rhoads:1999wm, Sari:1999mr, Harrison:1999hv}.	
As a consequence {{of the relativistic bulk motion}}, the flux is expected to be heavily reduced  if an observer's line of sight is not aligned within the jet. Therefore, there  {{may be}} less chances to detect prompt $\gamma$-ray emission from GRB jets directed away from us. However, Gamma Ray Bursts are followed by long lasting afterglow emission from $\gamma$-ray to radio frequencies \citep{Cota:1997cg, Frontera:1997ae, vanParadijs:1997wr, Frail:1997qf}. During the afterglow phase, the jet decelerates considerably and the doppler de-boost is alleviated making the emission visible to even observers  {oriented} away from the jet cone \citep{Moderski:1999ct, Rossi:2001pk, 2002ApJ...564..209D, Granot:2002za}. This makes afterglows to be potential candidates for electromagnetic (EM) follow-up observations  {{The joint event  GW170817 and GRB170817A was followed-up by various EM observatories \citep{MMApaper}. An optical/IR/UV transient consistent with predictions from kilonovae models \citep{Pian:2017gtc, Smartt:2017fuw, Arcavi:2017xiz} and X-ray/radio transient potentially from the GRB jet \citep{troja2017x, alexander2017electromagnetic, hallinan2017radio, Margutti:2017cjl} were detected.}}

%        (ADD CONTEXT, SAY THAT ADVANCED GW DETECTORS ARE COMING ONLINE. DESCRIBE THE DISCOVERY. ADD DETAILS ABOUT THE EM INSTRUMENTS. SAY THAT THERE COULD BE A POSSIBILITY THAT GW IS OBSERVED AND AG ARE OBSERVED IN VARIOUS WINDOWS. ADD A PARAGRAPH ON THE AG OBSERVATIONS (SEE RATES PAPER) WE STUDY VARIOUS POSSIBLE OBSERVATIONAL SCENARIOS OF JOINT DETECTION OF GW AND AG IN THE AG PARAMETER SPACE.)
        
  {With this observation, it became clear that neutron star mergers observed in the gravitational wave window may be accompanied with Gamma Ray Burst prompt emission and subsequent longer wavelength afterglows. Here, we focus on the joint detection of a GW event and the afterglow of the associated GRB.} In an associated paper \citep{saleem-etal-2017-agRates} we report the rate of afterglow detections in X-ray, optical, and radio wavelengths in such a scenario. 

%                Due to insufficient sky coverage of GRB instruments (ADD WHAT IS THE TYPICAL SKY COVERAGE ETC) as well as the short ($<2seconds$) duration, it is highly likely that we miss the prompt $\gamma$-ray emission from SGRB.                However, the afterglows(AG) of the SGRBs
%        which follow the prompt emission  are typically observed for days to weeks(months to years) in high(low) frequencies. The afterglow emission which is a synchrotron radiation powered by the interaction of the relativistic ejecta with the circumburst medium is governed by several (intrinsic) physical parameters pertaining to the source as well as the surrounding medium. The observed flux is further governed by additional observer dependent (extrinsic) parameters too. This together forms afterglow parameter space. 

        In this work, we  systematically explore the
        influence of the afterglow parameter space (particularly the
        ranges and distributions of the parameters) on the detectability of different
        afterglow components, for observations in
        coincidence with the GW-detected BNS merger events. We have
        considered EM facilities in X-ray, optical, and radio bands for afterglow detections. Similarly we considered a 
        5-detector network of GW detectors for BNS merger detections. { 
        We recall that the the first four GW detections from Binary Black-hole
        events \citep{gw150914,gw151226,gw170104,gw170608} were made by a
        2-detector advanced LIGO network and most recently, the first ever 3-detector observation of gravitational waves was reported from a BBH system with LIGO-Virgo network \citep{gw170814} followed by the BNS merger GW170817 \citep{GW170817}. With more detectors, we expect to observe more number of compact binary mergers.} Our studies are done with simulated BNS sources whose physical parameters are distributed typically within their ranges inferred from observations. 
        %{\red With possible
        %parameters used as proxies, we have associated the GW detectable
		%BNS mergers to SGRBs.} 
		With simulated afterglow
        light curves, we investigate various components of the afterglow parameter space
        which could influence detections, and we specifically focus on the {{observer's viewing angle with respect to the jet axis.}} We divide the afterglow population into two: \textit{within-jet} cases where the observer's line of sight {points to} within the jet cone, and \textit{outside-jet} cases where the line of sight falls outside the jet cone. With different detection/non-detection  scenarios in X-ray, optical, and radio bands, we identify favourable regions in the afterglow parameter space for both cases.

%ADD A PARAGRAPH OF THE SUMMARY OF RESULTS 
We observe that most within-jet afterglows are detected by X-ray and
optical instruments independent of other afterglow parameters.
However, only radio afterglows are expected to be detected from sources
where the observer line of sight is directed far off the jet edge. Even in such cases, afterglow parameters like jet energy and ambient medium density are critical for radio observations. 

        The paper is organized as follows.
        In section-2 of the the paper we give a description of the multi-dimensional afterglow parameter space, and explain the basic evolution of the afterglow spectrum and lightcurve.   Section-3 discusses the simulated SGRB population and their association  with GW-detectable BNS merger events. We present our results and findings in section-4 explaining how the afterglow parameter space results in different multi-band detection scenarios. We summarize our results in section-5. %{\red{need to polish this sentence. We present our results and findings in section-5 explaining the distribution of peak flux as a function of various components and distributions of parameter space components along with their implications as different detection scenarios.}} 

\section{Gamma Ray Burst afterglows}
    \label{sec-ag-theory}

	In this section, we discuss the basics of afterglow theory and
	introduce the multi-dimensional afterglow parameter space.

	GRB afterglow emission arises from the interaction of the jet with	the medium surrounding the burst \citep{1992MNRAS.258P..41R, Paczynski:1993gz}. The ultra-relativistic shock generated in the ambient medium enhances the magnetic field in the shock downstream and
	accelerates particles to high
	energies. The non-thermal electron population accelerated by the shock
 radiates {\it via} the synchrotron process. This radiation is seen as
	the afterglow emission. The shock decelerates as it encounters more
	material, therefore the thermal energy density it deposits in the
	downstream decreases with time \citep{Blandford:1976uq}, giving rise to a time evolving
	afterglow light curve in frequencies ranging from $\gamma$-rays to
	radio \citep{1993ApJM, Meszaros:1996sv, 1997ApJ...489L..37S}. The GRB afterglow emission peaks in high frequencies first,
	followed by lower frequencies. Typically, afterglows can be observed
	for several days in X-ray/optical to months and years in radio. See \cite{1999PhR...314..575P, 2006RPPh...69.2259M, 2009ARA&A..47..567G, Kumar:2014upa} for reviews.

{{In addition to the above described forward shock which is moving in to the external medium, a reverse shock travels back to the ejecta which can produce bright early afterglow emission especially in optical and radio bands \citep{1993ApJM, Akerlof:1999aa, Sari:1999kj}. We have not considered this component in the paper. }}
%For almost
%74\% of the detected SGRBs, afterglows in X-ray have been
%observed. Optical has been observed for 34\% and radio only for 7\%
%\citep{fong2015decade}.
%
%Due to strong relativistic beaming, GRB prompt emission is expected to be detected only if the observer's line-of-sight lies within the cone of the jet. Therefore, even when an NS merger produces a relativistic jet, the GRB can be missed if the jet is not directed towards the observer. 

\subsection{Afterglow parameter space}
%{\red{the intrinsic vs extrinsic is foreign to grb theory. kuntal and myself discussed and decided to rewrite it mildly.}}

There are six physical parameters intrinsic to the emitting plasma that decide the afterglow
spectral evolution with time. These are; the isotropic equivalent kinetic energy $E_{\rm
  iso}$ carried by the jet, initial half opening angle of the jet
$\tj$, number density $n$ of the circumburst medium (assumed to be
homogeneous), fraction $\eB$ and $\eE$ of the shock thermal energy in
downstream magnetic field and non-thermal electrons respectively, and
power-law index $\p$ of energy-spectrum of the non-thermal electrons
radiating synchrotron emission.  We are not considering dust and gas
absorption due to the intervening medium.

Apart from these six parameters, there are two parameters external to the emitting region, distance $D_L$ and the angle $\tv$ between the observer's line of sight and jet axis. Unlike in gravitational wave studies, $D_L$ is a fixed parameter as it is known sufficiently precisely (albeit with an underlying cosmological model) through redshift from optical spectroscopy in most cases. Therefore, the final afterglow parameter space is a $7$-dimensional one.
%These parameters together constitutes the $7$-dimensional afterglow parameter space.

In the next section, we describe how these parameters enter in the
expression of the afterglow flux evolution ($f_{\nu} (t)$) measured at
the observed frequency $\nu$ and at an observer time ($t$) measured 
from the GRB trigger time.

\subsection{Afterglow spectral evolution and lightcurves}

\begin{figure}
	\includegraphics[scale=0.45]{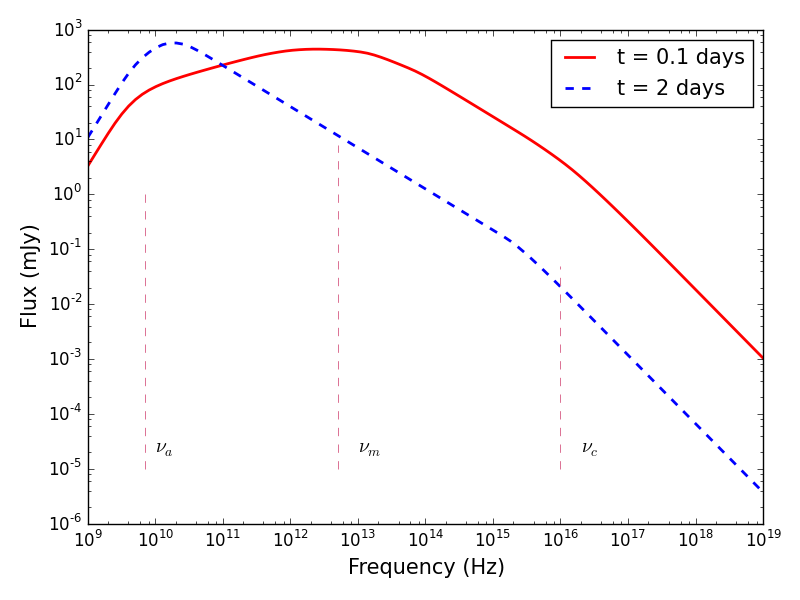}
	\caption{ Afterglow spectrum for $t=0.1$~day since burst (red). The parameters used are $E_{\rm iso} = 10^{51}$~erg, $n =
          1.0$~atom/cc, $\tj = 5^{\circ}, \eE = 0.1, \eB = 0.01$, and
          $p = 2.5$. The observer is on the axis of the jet at
          $300$~Mpc away. Locations of the break-frequencies $\nu_c,
          \nu_a,$ and $\nu_m$ are marked. The X-ray band
          is above $\nu_c$, optical frequencies are between $\nu_m$
          and $\nu_c$, high radio frequencies are below $\nu_m$, while
          low radio frequencies are below $\nu_a$ (optically
          thick). The blue curve shows the evolution of the spectrum at a later epoch ($t=2$~day).  The order of these frequencies can change depending on the physical parameters. For example, higher number densities can result $\nu_a  > \nu_m$  and for high magnetic field values, i.e., for larger values of $\eB$, $\nu_c$ can be below $\nu_m$.}
	\label{agspec}
\end{figure}

Synchrotron spectrum of a single electron peaks at a characteristic
frequency ($\nu_{\rm syn}$) governed by the average magnetic field
($B$) of the plasma, and the lorentz factor ($\gamma$) of the
electron.  Afterglow synchrotron spectrum at a given epoch from a collection of
electrons in the 
power-law distribution is obtained by a convolution of the
single electron spectrum and the electron distribution \citep{rybicki1979radiative}. 
It can be
approximated as a combination of piecewise power-law segments
separated by three break-frequencies \citep{ Sari:1997qe, Wijers:1998st}; (see Fig.\ref{agspec}).
These break-frequencies are: (i)
$\nu_c$, the frequency above which synchrotron radiative losses are
severe; (ii) $\nu_a$, the frequency below which the fireball is
optically thick; and (iii) $\nu_m$, the characteristic synchrotron
frequency of the lowest energy electron in the plasma (corresponding
electron Lorentz factor $\gamma_m$). As the injected electron distribution is a power-law, the number of electrons will be the highest at $\gamma_m$. Therefore, the synchrotron spectrum will peak at $\nu_m$.  
The spectral peak, $f_m$ is the fourth and
final spectral parameter, which is defined as the flux
at $\nu_m$. However, in the presence of intense synchrotron losses, electron distribution will evolve to lorentz factors below $\gamma_m$. Or in other words, $\nu_c$
falls below $\nu_m$ and the spectral peak shifts to $\nu_c$ \citep{Sari:1996za, Waxman:1997ga}. 

%$\nu_c, \nu_m, \nu_a, f_m$ 
As the slope of all the spectral segments will be uniquely determined by the power-law
index $p$, the four spectral parameters described above and $p$ together determines the flux $f_{\nu} (t)$.  These four spectral parameters are functions of the physical parameters $E_{\rm iso}, n, \eE, \eB$. %In addition, the spectral peak $f_m$ (which can also be seen as the flux normalization) depends on the luminosity distance $D_L$. %
Therefore, in general the five physical parameters ($E_{\rm iso}, n, \eE, \eB, p$) uniquely decide
$f_{\nu}$ at a given $t$.  However, in different synchrotron spectral regimes, the dependency will be different. For example, for a fixed index of $p=2$, in $\nu_m < \nu < \nu_c$, $f_{\nu} \propto {E_{\rm  iso}}^{3/2} \, \eE\, \eB \, n^{1/2}$. If the observed frequency is above $\nu_c$, flux is independent of $n$
and $f_{\nu} \propto E_{\rm  iso} \, \eE\, {\eB}^{-1/4}$.  The dependencies are
more complex in the optically thick regime. See \citep{Wijers:1998st} for details. In the example, shown in figure-\ref{agspec}, the spectrum is calculated at $t=0.1$~day since burst for a typical parameter set. Here, the X-ray frequency is above $\nu_c$, optical is between $\nu_m$
and $\nu_c$, and radio is below $\nu_m$ or $\nu_a$. 

%Below, we describe dependence of fluxes in different spectral
%segments. Typically this dependence can have on different physical
%parameters. 
%
\subsubsection{Shock dynamics and lightcurves}
\label{onaxislc}
%(CAN WE MAKE 2.2.1 AS SHOCK DYNAMICS PARAMETERS? 2.2.2 lIGHTCURVES FOR ON-AXIS OBSERVERS AND 2.2.3 LIGHTCURVES FOR OFF-AXIS OBSERVERS?)
%The afterglow light curve carries information of the afterglow dynamics.
To obtain the afterglow light curve, the evolution of $\nu_c, \nu_a, \nu_m$, and $f_m$
needs to be calculated ($p$ is assumed to be a constant over
time). Time evolution of spectral parameters is a consequence of the
afterglow dynamics, i.e., the evolution of the bulk lorentz factor
$\Gamma$ and the radius $R$ of the shock front, which in turn is determined by $E_{\rm iso}, n,$ and $\tj$. Therefore, with $6$ physical parameters (i.e., $E_{\rm iso}, n, \eE, \eB, p, \tj$),
afterglow flux evolution $f_{\nu} (t)$ can be calculated for an
observer along the axis of the jet \citep{Panaitescu:2000xk, Panaitescu:2001fv, Resmi:2008qb}.

%(2.2.2 CAN START HERE)\\
\subsubsection{Lightcurves for on-axis observers}
As the spectrum evolves with time, a given observed frequency moves across different spectral segments. For on-axis observers (i.e., for $\tv = 0$), if the observed frequency is in the optically thin part of the synchrotron spectrum (i.e, above $\nu_a$), the light curve peaks when its frequency crosses $\nu_m$. For low frequencies like radio, which are likely to be below $\nu_a$ (i.e, for which the fireball is optically thick) the light curve peak delays till $\nu_a$ crosses the band \citep{Panaitescu:2000bk, Resmi:2005bj}.

%Light curve plot
\begin{figure*}
	\includegraphics[scale=0.45]{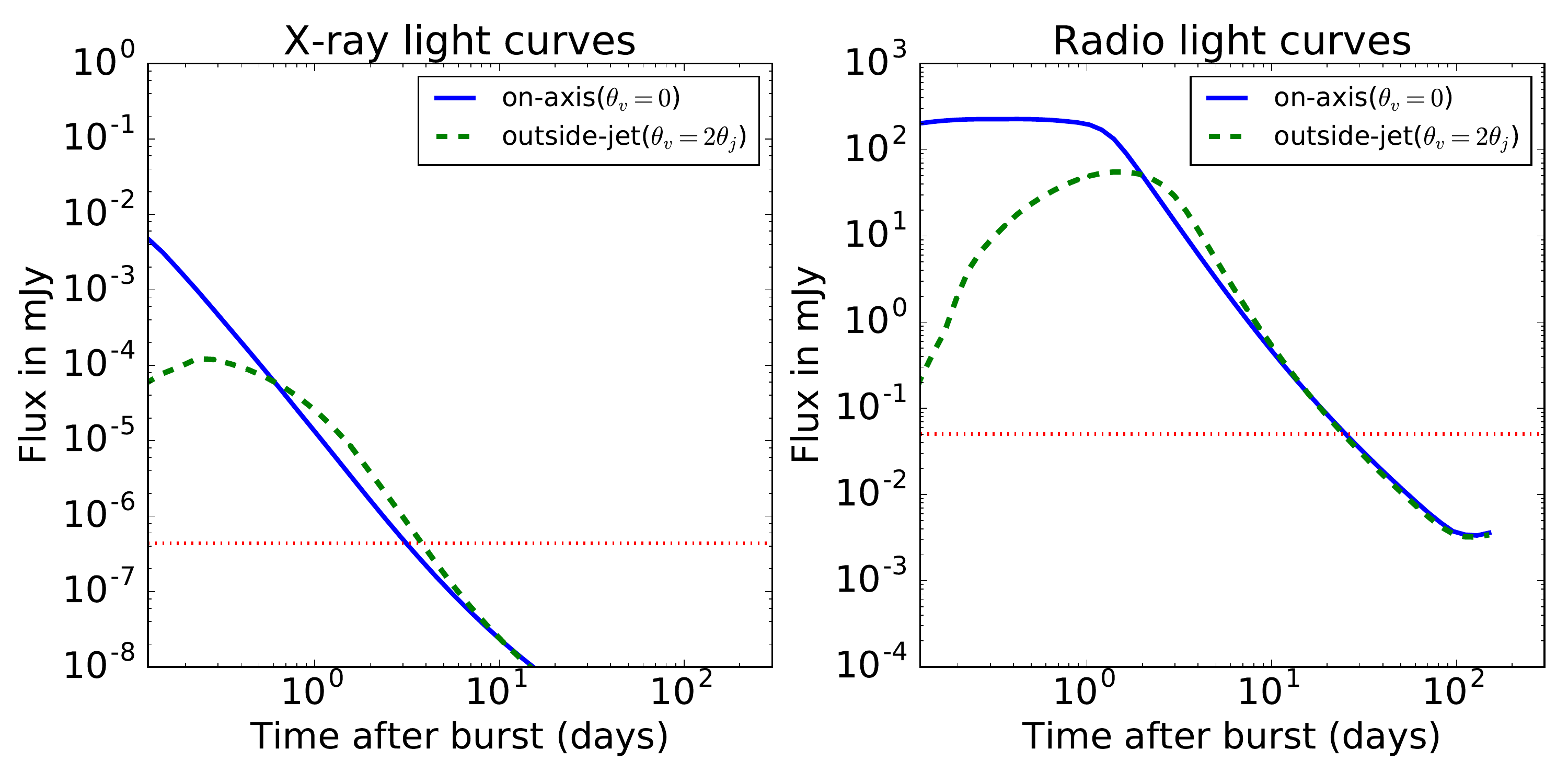}
	\caption{ X-ray and radio afterglow light curves for on-axis and off-axis (as $\tv/\tj >>1$, this is an outside-jet case) observers computed using the \textit{BoxFit} code. The X-ray peak corresponds broadly to $t_{\tv}$ (see text). The radio peak is delayed due to self-absorption effects. The off-axis lightcurve goes above the on-axis one because of photons from such latitudes that require a longer light travel time. At a given observer time, these photons correspond to an earlier jet co-moving frame time when the emissivity was higher.}
	\label{lcfig}
\end{figure*}

\subsubsection{Lightcurves for off-axis observers}
\label{offaxislc}
For off-axis observers, $\tv$, the viewing angle, enters the picture through relativistic effects. Due to the high lorentz factor of the jet, flux observed at line-of-sights which are off the jet-axis will be severely doppler de-boosted. The de-boost is relaxed at  $t_{\tv}$, when the monotonically decreasing $\Gamma$ goes below $1/\tv$ \citep{Moderski:1999ct, Granot:2002za}. Optically thin frequencies, like X-ray, will peak at $t_{\tv}$. In low radio frequencies, the peak arrives at a later epoch when the fireball becomes optically thin. See figure-\ref{lcfig} for X-ray and radio lightcurves for on-axis and off-axis ($\tv = 2\tj $) observers. The parameters used are same as that of figure-\ref{agspec}, except for the value of $\tv$. Around $t_{\tv}$, off-axis observer starts to receive photons from areas of the jet surface that was so far invisible due to relativistic beaming. The photons have a longer travel time compared to the on-axis observer, therefore at a given observer time the off-axis observer receives photons emitted at an earlier time compared to those received by the on-axis observer. This results in the slight increase of off-axis lightcurve post peak \citep{Granot:2002za}. 

\subsection{\textit{BoxFit} numerical hydrodynamic code}
To compute the evolution of the afterglow lightcurves, we use \textit{BoxFit}, a numerical hydrodynamical code which follows the evolution of the shock from ultra-relativistic to non-relativistic dynamics and calculates the afterglow synchrotron spectrum considering full relativistic effects \citep{vanEerten:2009pa, vanEerten:2010zh}. \textit{BoxFit} assumes a uniform top-hat jet model.  {{These calculations do not consider additional energy input from the central engine during the early afterglow phase, which is observed in some \textit{Swift} short bursts \citep{Campana:2006ja, Grupe:2006uc}.}}
%ADD REFERENCE TO BOX FIT. SOME WORKS UNDER BOX FIT. ANY LIMITATIONS OF BOX-FIT.
%=================================================================================        

   \section{Population of GW and SGRB afterglow events}
   Our focus in this study is to explore various features and properties of the afterglow lightcurves in X-ray, optical and radio bands detected as counterparts to the GW-detected BNS merger events. For our study, we use the simulated populations of SGRB afterglow lightcurves where each one is
   characterised by a set of $7$ afterglow 
   parameters which we discussed in detail in section.\ref{sec-ag-theory}.
   
   \subsection{Associating afterglows with the GW source}
   \label{sec-gw-ag-asso}
   As mentioned earlier, it is believed that the BNS mergers
   which produce gravitational waves also power short-GRBs
   followed by its associated afterglows in various EM bands.
   Amongst the afterglow parameters described in
   section.\ref{sec-ag-theory}, the luminosity distance $D_L$ and
   viewing angle $\tv$ are also essential to characterise the GW
   signals produced by the binary merger event. Note that viewing angle
   $\tv$ is referred to as the inclination angle $(\iota)$ of the
   binary which measures the angle between the angular momentum vector and
   the observer's line of sight in GW literature.
   
   Since we are considering afterglow events observed as
   counterparts to the observed BNS events, we need parameters to
   associate the two observed phenomena to one physical
   origin. In a realistic observation scenario, the primary key for association
   is the spatial and temporal coincidence of the GW-SGRB events. Practically, it is challenging to establish temporal coincidence in cases where there is no prompt emission detected.
   However, for our simulated population of GW-SGRB joint events, we assume that they do have temporal as well as spatial coincidence. As discussed above, binary inclination or the viewing angle $\theta_v$ as well as $D_L$ are common parameters for both GW and SGRB and hence they are the obvious choices for the GW-SGRB association in our simulated joint events.   
   In future, it might be possible to have a  much tighter association when more
   parameters can be identified in common for describing both phenomena.
   For example, the burst properties such as disk mass $M_{\rm disk}$ or ejecta mass $M_{\rm ejecta}$ can be constrained from the inference of intrinsic binary parameters such as component masses,
   spins and the equation of state parameters \citep{giacomazzo2012compact,foucart2012black,kawaguchi2016,dietrich2017modeling} and this in turn may be used to place bounds on the isotropic energy $E_{\rm iso}$
   
   In realistic observations, the error on gravitational wave inferred
   distance can be reduced if the red-shift measurement $z$ can be obtained
   from the spectroscopy of either the galaxy associated with the GW event or an optical
   counterpart (afterglow or kilonova). The distance estimates can in turn
   improve our knowledge of inclination of the source~\citep{arun2014synergy}

   \subsection{GW-detected BNS mergers}
   \label{sec-gw-sim}
   Here, we assume that the complete network of ground based advanced
   detectors is functioning. Thus, we consider a 5-detector network
   LHVKI which includes  LIGO-Livingstone(L), LIGO-Hanford(H), 
   Virgo(V) and the two upcoming detectors Kagra(K) and LIGO-India(I). 
   For convenience, we assume that all the detectors have achieved similar sensitivity
   as that of LIGO's designed sensitivity. 
%============================================
\begin{figure}
	\includegraphics[scale=0.4]{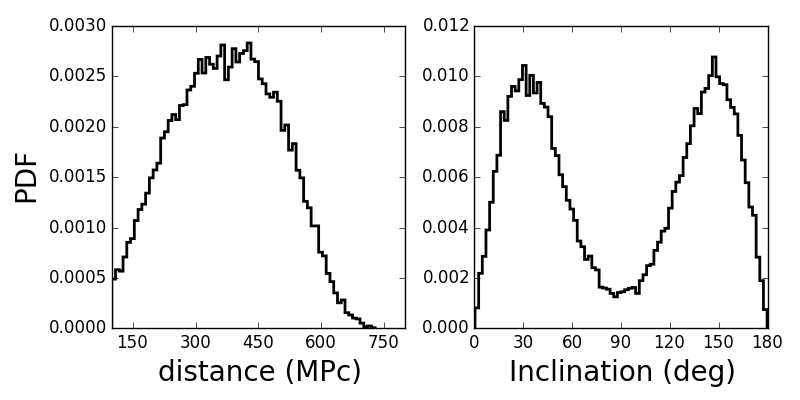}
	\vspace{-3mm}
	\caption{Distributions for distance and inclination ($D_L\&\iota$) of non-spinning BNS sources detected at LHVKI network with a detection criterion of minimum network SNR of 8. This can be used as representative distributions of sources considered for EM follow up observations in the 5-detector era of advanced ground based GW detectors.} 
	\label{BNS-pop}
\end{figure}

   We simulate $3\times10^5$ non-spinning BNS sources with component masses 1.4$\Msun$ each,  uniformly distributed in comoving volume between 100-740Mpc. Inclination($\tv$ or $\iota$) of the population is distributed as uniform in $\cos\iota$. For each source, we simulate GW signal using the analytical 3.5 order post-Newtonian TaylorF2 waveform \citep{Bliving,BDEI04,BDIWW95} and compute the network SNR(Signal-to-Noise Ratio)~\citep{Pai2000}. We consider all sources with minimum network SNR of 8 as being detected. Applying this criterion, we obtain $\sim5\times10^4$ sources with LHVKI up to a  maximum distance $\sim730Mpc$. 
   
   Figure.\ref{BNS-pop} shows the distribution of distance($D_L$) and
   inclination ($\iota$) of the detected BNS sources. For the actual
   simulated population which is uniformly distributed in volume, the distribution of distance follows $P(D_L) \propto D_L^2$ such that larger the distance more the number of sources. In addition, due to the antenna pattern
   functions, the sensitivity of GW detector networks is not isotropic, rather varies for different directions.  Hence the sources located at highly sensitive
   regions in sky can be detected even if they are deep in distance whereas only nearby sources can be detected from less sensitive regions in sky. 
   On an average, this results in less number of detectable sources at
   larger distances as shown in the left panel of Figure.\ref{BNS-pop}. Similarly, in the simulated population, $\iota$    is distributed as uniform in $\cos{\iota}$, (ie, $P(\cos{\iota})  \propto {\cal{U}}(-1,1)$ which translates as $P(\iota) \propto
   \sin{\iota}$). However, as seen in the right panel of Figure.\ref{BNS-pop}, the
   inclination distribution of detected sources is biased towards
   face-on sources ($\iota\rightarrow 0$ or $180$~degree). Thus, face-on
   sources are detectable to much larger distances than the edge-on
   sources ($\iota\rightarrow 90$~degree). We have drawn 50,000 BNS sources from the
   detections and associated them to the 50,000 SGRB sources
   in population-1\&2 described below.

   \begin{table}
   	\centering	
   	\label{tab:param-distr} 
   	\caption {Components of afterglow parameter space along with their ranges and distributions.  Prior range and distribution for the $D_L-\iota$ combination which are observer dependent are obtained from a GW-detectable distribution of BNS sources as shown in Figure.\ref{BNS-pop}. Remaining parameters are intrinsic to the afterglow generating mechanism and their prior ranges are taken as inferred by observations and theory. Their distributions are not well known and we have considered two types of distributions labelled as population-1 and population-2[Please see text for more details].}
   	
   	\begin{tabular}{lccr}
   		\hline
   		\textbf{Parameter}& \textbf{Range} & \textbf{Population-1} & \textbf{Population-2 }  \\
   		\hline	
   		$D_L$	       & --          & GW prior	              & GW prior	      \\
   		$\tv$	       & --          & GW prior	              & GW prior	      \\
   		$\tj$          & $(3\dg,30\dg)$	  & $P(\tj)\propto \cal{U}$ &$P(\tj) \propto \cal{U}$ \\
   		$\Eiso$(erg)&$10^{49}-10^{52}$ &$P(\log{\Eiso})\propto \cal{U}$&$P({\Eiso})\propto \cal{U}$ \\ 
   		$n$	($cm^{-3})$& 0.0001-0.1  & $P(\log{n})\propto \cal{U}$  & $P(n) \propto \cal{U}$  \\ 
   		$\eB$          & 0.01 - 0.1  & $P(\log{\eB})\propto \cal{U}$& $P(\eB)\propto \cal{U}$ \\ 
   		$\eE$	       & 0.1         & fixed                  & fixed             \\
   		$p$            & 2.5         & fixed	              & fixed  \\												
   		\hline
   	\end{tabular}
   \end{table}

   \subsection{Short-GRB population choices}	
   We have two populations namely population-1 and population-2, with
   each of them containing 50,000 SGRB afterglow events(sources). The
   two populations differ in the choice of priors on parameters
   $E_{iso},n,\eB$. For both the populations, we consider $\Eiso$
   ranging in the typical limits between $10^{49}$ erg and $10^{52}$
   erg, number density between $10^{-4}-10^{-1}$ and energy fraction
   in the magnetic field $\eB$ ranges between $10^{-2}-10^{-1}$.  In
   population-1, we have drawn $\Eiso, n$ and $\eB$ within the ranges described above as uniform in  $\log{\Eiso}, \log{n} $ and $\log{\eB}$ respectively and in
   population-2, as uniform in ${\Eiso}, {n} $ and ${\eB}$ respectively. The two types of
   priors are the two limiting distributions we consider here and 
   expect that they would capture the essence of variations in distributions.  
   Moreover, the log prior in energy also reflects the luminosity function of
   short GRBs, which is believed to be of a power-law nature
   \citep{Guetta:2004fc}. Log priors in $\Eiso$ and
   $n$ ensure that there is a considerable number of bursts with lower
   energies and number densities. The jet half opening angle $\tj$ is
   uniformly distributed between $3-30$ degrees for both the
   populations. We fix the upper limit to $30$ degrees following the values of jet collimation angles from a
numerical simulation of binary NS merger by \cite{rezzolla2011missing}. We fixed $\eE$ and $p$ at fiducial typical values of $0.1$ and $2.5$ respectively. While the flux is not highly sensitive to the value of $p$, radio afterglow observations indicate that $\eE$ is confined to a narrow range \citep{Beniamini:2017jil}.
   
   The remaining two parameters $D_L$ and $\tv$ are the ones which are
   used to associate the afterglows and GW signal, as discussed in
   section.\ref{sec-gw-ag-asso}. Since our population should be
   representative of EM follow-up observations of BNS merger
   detections, we draw these two parameters from a simulated
   distribution of GW-detectable BNS population as discussed in
   section \ref{sec-gw-sim}. The complete set of parameters along with the
   priors are given in Table.\ref{tab:param-distr}.

	\begin{figure*}
		\includegraphics[scale=0.34]{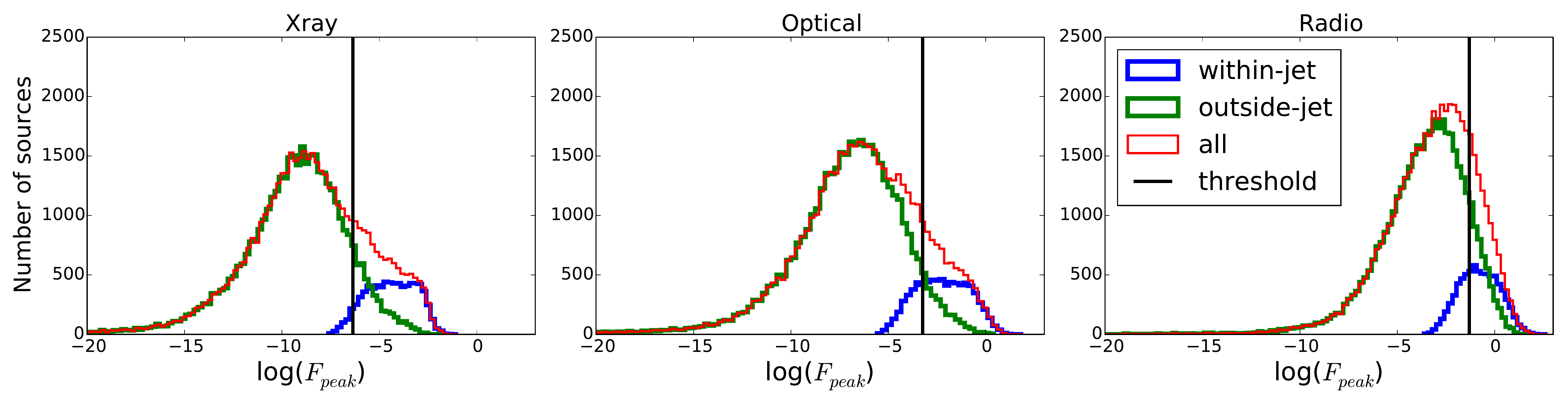}
		\caption{Afterglow peak Flux distributions in X-ray, Optical and Radio bands for 50,000 sources detected in GW detector network LHVKI. The black vertical lines in each panel are the detection thresholds of \textit{XRT}, \textit{LSST} and \textit{JVLA} respectively. X-ray and optical peak flux distributions are made up of two bell-curves. The smaller one at the right are for within-jet sources (i.e,$\tv<\tj$) and the larger one at the left are outside-jet sources ($\tv>\tj$) . Radio has a symmetric peak-flux distribution due to reduction in doppler beaming at the time of typical radio peaks (see text for a detailed description). These figures are made for population-1. Please note that the legends in 3rd panel apply to all 3 panels.}
		\label{fig-Fpeak-distr}
	\end{figure*}
\begin{figure*}
	\includegraphics[scale=0.34]{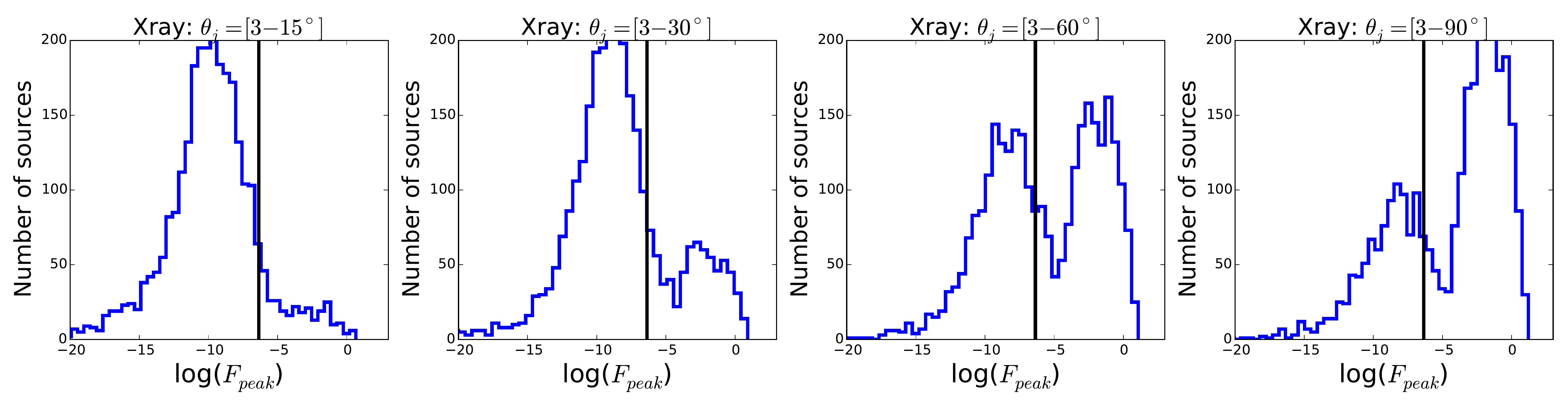}
	\caption{Afterglow peak flux distributions in X-ray for populations with different ranges of $\tj$ where other parameters are distributed as in population-1. Given the $\tv$ distribution, for smaller ranges of $\tj$ (for eg, left panel where $\tj:3-15\dg$ ), less sources are within-jet and majority are outside-jet whereas for larger ranges of $\tj$(for eg, right panel where $\tj:3-90\dg$ ) relatively more sources are within-jet. This reflects in the statistical properties of the two bell shaped curves in each panel.}
	\label{fig-tj-effects}
\end{figure*}

%==========================================================
	
	%==============================================		
	\begin{figure*}
		\includegraphics[scale=0.34]{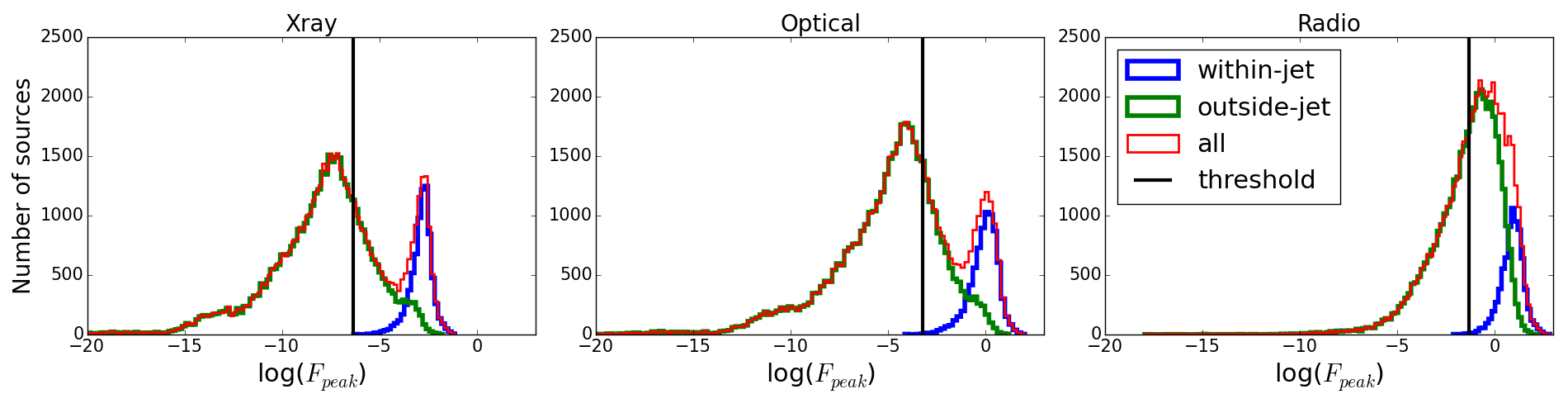}
		\caption{Same as figure-\ref{fig-Fpeak-distr} but for population-2. Changes in the distribution of $E_{\rm iso}$, $n$, and $\eB$ (though the ranges of these parameters are the same) has influenced the distribution of peak fluxes. See text for details.}
		\label{fig-Fpeak-distr2}
	\end{figure*}
	%===============================================
%==========================================================	
\section{Simulation Results}
For all the SGRB sources in population-1 and 2, we carry out \textit{BoxFit}
simulations and compute the afterglow light curves in X-ray, optical and radio frequencies. Specifically, we compute lightcurves at frequencies $2.4 \times 10^{18}, 4.5 \times 10^{14},$ and $15 \times 10^{9}$~Hz for X-ray, optical, and radio bands  respectively.

We consider three representative instruments, the \textit{Swift}-XRT
in X-ray, the Large Synoptic Survey Telescope (LSST) in optical
(R-band), and the Jansky Very Large Array (JVLA) in radio. The XRT
sensitivity of $10^{-14}$~erg cm$^{-2}$ sec$^{-1}$ for $10^{4}$~sec integration in $0.3 - 10$~keV
\footnote{https://swift.gsfc.nasa.gov/about\_swift/xrt\_desc.html} is
converted to mJy by assuming a flat spectrum. The corresponding XRT
threshold at $2.4 \times 10^{18}$~Hz comes out to be $4.37 \times
10^{-7}$~mJy. We adopt a single visit R-band sensitivity of $24.5$
AB-magnitude for LSST. The corresponding detection threshold will be $5.75 \times 10^{-4}$~mJy \footnote{$f_{\nu} = 10^{-\frac{m_{AB}-16.4}{2.5}}$~mJy}. We considered $50 \mu$Jy to be the $3
\sigma$ limiting flux required for radio detections at $15$~GHz.

If at least one point in the simulated lightcurve is above the
threshold, we consider it as a detection. Since the lightcurves are
sampled logarithmically ($\delta t/t \sim 1$), this condition is
sufficiently conservative. We use
$5$~hours since burst as the start time of the observations. This
particularly ensures that the above mentioned detection criteria is consistent with the $10^{4}$~sec integration time required for XRT. {{In addition, this criteria also helps us stay clear of the early afterglow phase which could be influenced by delayed flares from the central engine.}}
However, in the  companion paper \citep{saleem-etal-2017-agRates},
we demonstrate how cadence can affect the detection statistics.
 
Below, we summarize our results and findings of the synthetic joint observations of GW from BNS and the associated afterglow detection in the EM window.  We discuss
various plausible joint observation scenarios for the same.

\subsection{Peak flux distribution} 

As discussed above, if at least one point in the simulated
lightcurve is above the threshold, we call it a detection. In other
words, if the peak flux of a given lightcurve is below the threshold,
it will not get detected. Therefore, the distribution of peak flux is
a proxy to understand the influence of the GRB parameter space on
detection.

We compute the peak flux for the 50,000 simulated SGRB lightcurves  (counterparts
to GW-detected BNS mergers). Defining $\tvj \equiv \tv/\tj$, first of all we divide the afterglows in two cases: 1. {\it within-jet} ($\tvj < 1$) cases where the observer's line of sight aligned within the jet cone and 2. {\it outside-jet} ($\tvj > 1$) cases where the line of sight directs somewhere off the jet cone. We investigate the lightcurve behaviour for these two cases separately. Figure-\ref{fig-Fpeak-distr} shows the peak flux
distributions in X-ray, optical and radio bands from population-1. The black vertical
lines in each panel are the detection thresholds considered for \textit{XRT},
\textit{LSST} and \textit{JVLA}.

\subsubsection{Characteristic features}
The peak flux distribution shown in Figure-\ref{fig-Fpeak-distr} is a
combination of two bell-shaped curves, corresponding to the
within-jet and outside-jet cases together.
The area under the blue(green) curve represents the total
number of within(outside)-jet cases and the red curve shows the entire sample.  We see clearly that the total number of within-jet cases is smaller than that of
the outside-jet cases.
 This is a reflection of our original simulated afterglow sample
(only 15\% are within-jet cases) and is \textit{not} a consequence of
any detection criterion. Typically, {\it within-jet} cases have
higher flux as the doppler boost enhances their flux as opposed to the {\it outside-jet}
cases which are heavily de-boosted.

\begin{figure*}
	\includegraphics[scale=0.21]{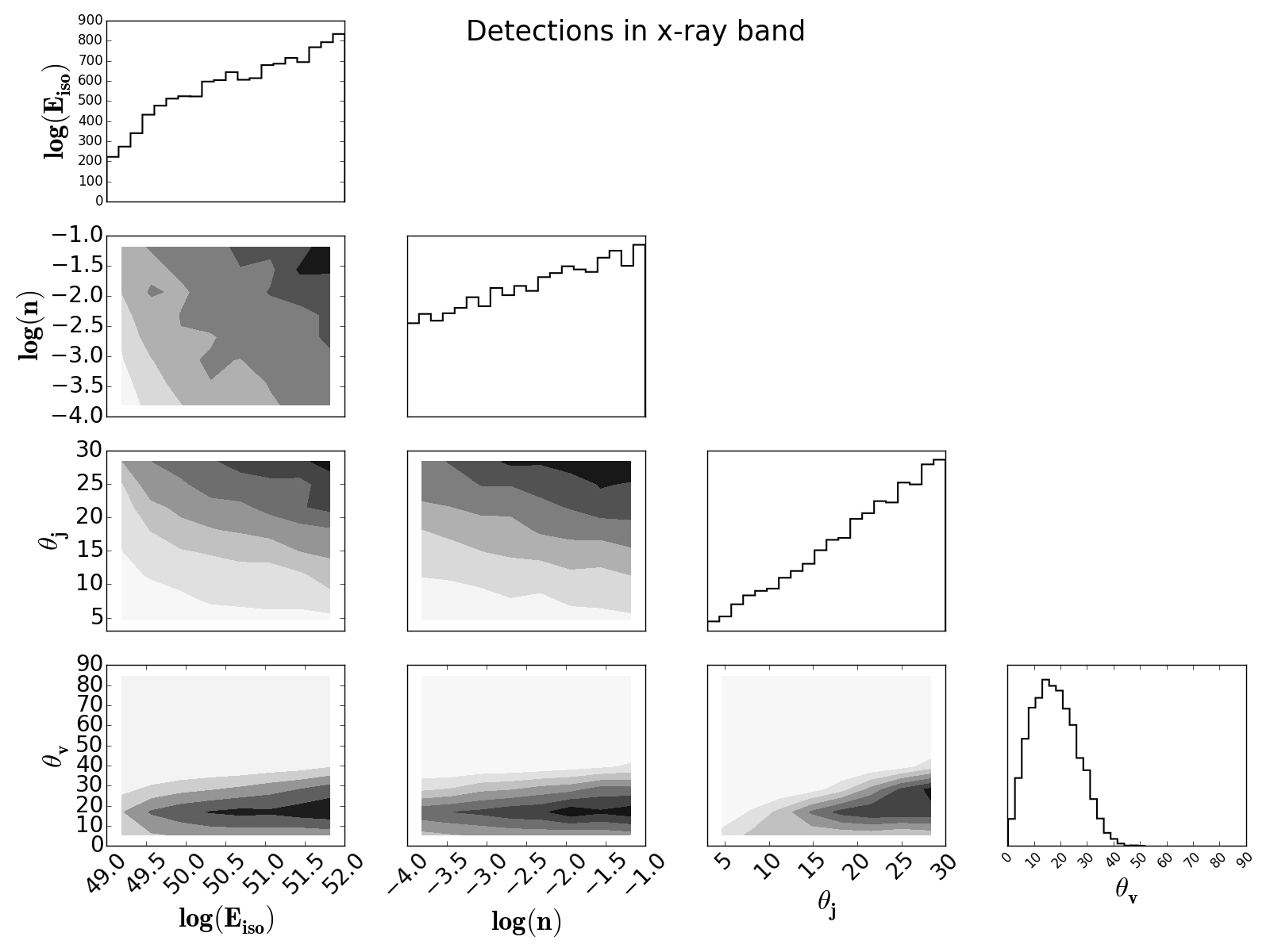}
	\includegraphics[scale=0.21]{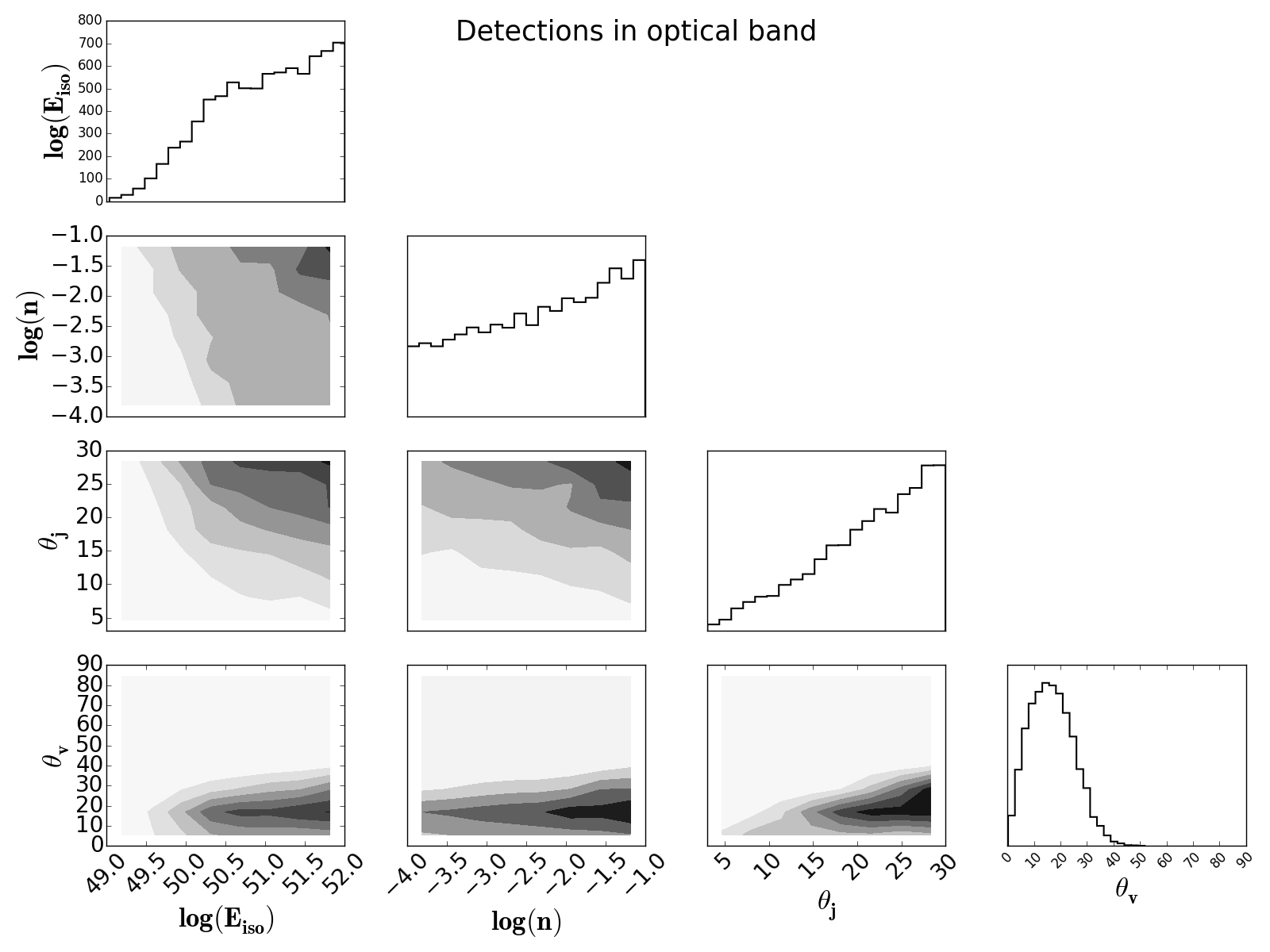}
	\hrule
	\includegraphics[scale=0.21]{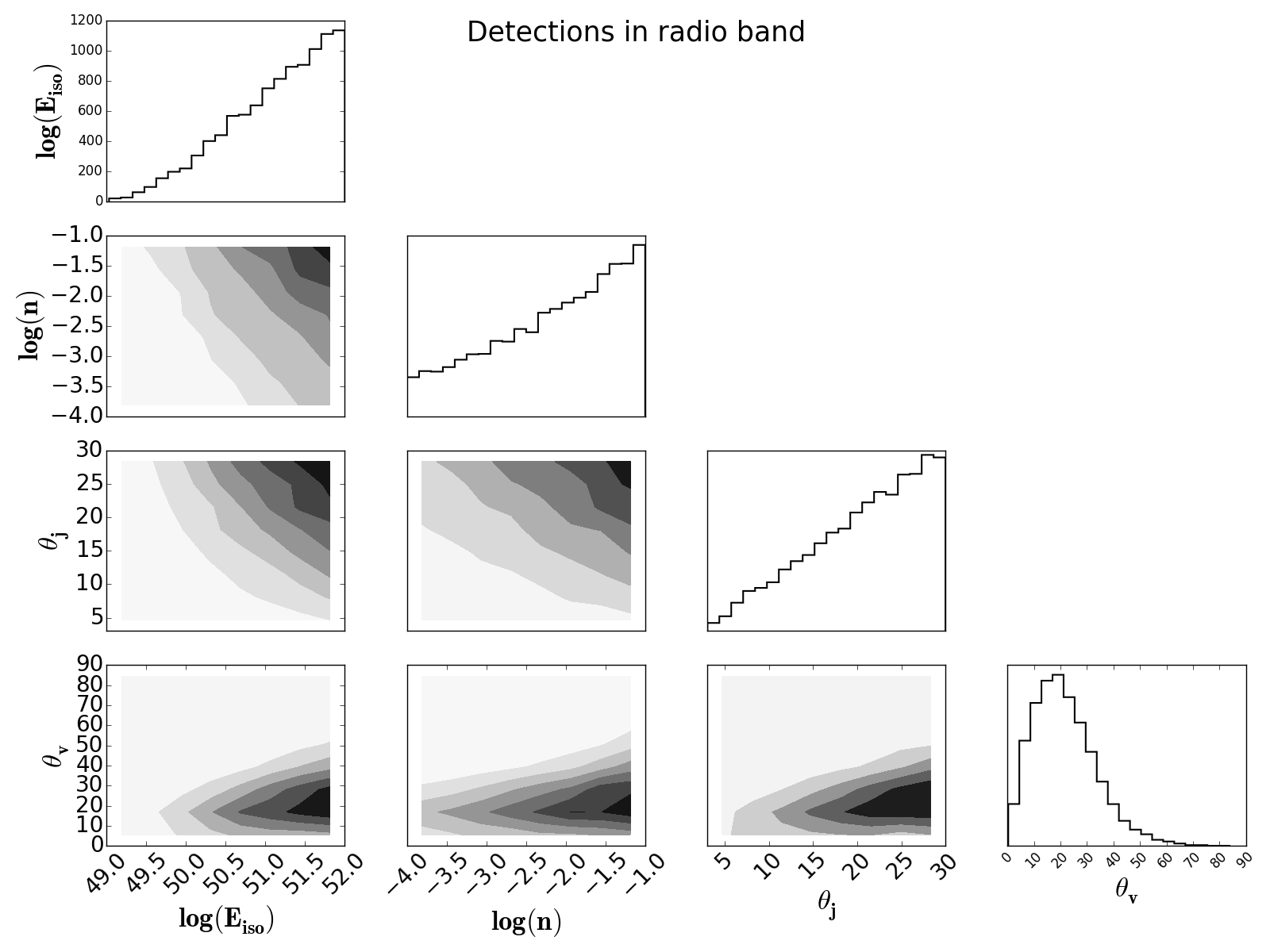}
	\caption{Corner plots showing the distribution of detected sources in X-ray(top left), optical(top right) and radio(bottom ) for population-1.  Each corner plot contains multiple panels showing the 2D histograms of different sub-spaces of afterglow parameters $E_{iso},n,\tj, \tv$ with the panels along the diagonal showing their histograms. In the 2D histograms, the dark region indicates higher probability. Though we have considered a range of $0.01< \eB < 0.1$, we have not included $\eB$ in the plots.}
	\label{fig-corner-XOR-all}
\end{figure*}

The {\it within-jet} and {\it outside-jet} cases 
appear prominently distinct
in the combined histogram (red) of the X-ray and the optical afterglow observations while
the radio show a smooth resultant distribution. This is because the
radio peaks are delayed compared to higher frequencies and the doppler
de-boost is well relaxed by their peak time (see section-\ref{offaxislc}).

\subsubsection{Radio vs. higher frequencies}
\label{rad-vs-higher}
%(AP:TRY TO CONNECT TO THE TYPICAL LIGHT CURVE FIGURE)

As explained in section-\ref{onaxislc}, the peak time $t_{\rm peak}$ (time at which the light curve peaks) for an on-axis ($\tv = 0$) burst is decided solely by the physical
parameters $E_{\rm iso}, n, \eB, \eE$. For $15$~GHz radio band,
this ranges from a few hours to a few days while X-ray and optical lightcurve peaks are already in
decline for the $\ts$ we use ($5$~hrs since burst). See the example figure-\ref{lcfig}. As mentioned in section-\ref{offaxislc}, for outside-jet cases, the peak is delayed till $t_{\tv}$.%$t_{\tv}$ increases with $\tv$. 
As can be seen from figure-\ref{lcfig}, for a typical radio lightcurve, $\tp (\tv=0) > \ts$ and in both $\tp  (\tv=0) $ and $t_{\tv}$ are roughly of the same order. Therefore, the doppler de-boost is not as severe in radio as it is in the higher frequencies. Moreover, the peak flux of within-jet cases and outside-jet cases therefore are not very different from each other in radio, leading to smooth peak flux distribution we observe in the
figures. %We have checked the $\tp$ of within-jet and outside-jet cases separately for all bands and found that this indeed is the case.

%(AP:MOVE THIS COMMENT TO RATES PAPER. The same effect is seen as `plateaus' in the curves of afterglow detection fraction in the companion paper on rates.)
	
It has to be noted that for the detection thresholds we used, in within-jet afterglows, the detection fraction in
X-ray and optical are higher than that in radio (area of the blue curve right of the black
vertical line). This is because of two factors: (i) the X-ray and optical thresholds are deeper than radio; and
(ii) the flux in both these bands are enhanced by doppler boost at their peak as opposed to radio which peaks later when the Lorentz factor and boost are relatively of lesser magnitude.

\subsubsection{Factors affecting the peak flux distribution}
We note that the peak flux distribution follows the bell curve for within jet and outside jet cases. Here, we investigate the factors affecting the properties of the bell curve. Each bell curve can be broadly approximated as a Gaussian. Thus, we can define a mean value ($x$), width($\sigma$), and height ($y$) of the curve.

Obviously, the height of the curves are strongly sensitive to the value of $\tj$ for a given distribution of $\tv$. Currently we have only 15\% of within-jet cases and this is because of the fact that we have let $\tj$ vary upto $30^{\circ}$. If $\tj$ is limited to a much lesser value around $10-15 ^{\circ}$, typically considered in the literature, we will have even lesser number of within-jet cases leading to a reduction in $y$ of the bell-curve of the right side. Instead, if we consider $\tj$ upto larger angles such as $60^{\circ}$ or $90^{\circ}$,  fraction of within-jet cases as well as the height of the right bell-curve will increase. This is an obvious effect, however is illustrated in Figure -{\ref{fig-tj-effects}} by taking X-ray afterglows as an example.
%This %in-turn will also influence the position of the plateau in
%fig-2 of paper-I. %. As the height of the distribution changes, the
%dip(part where two distributions meet) also changes horizontally.

\begin{figure*}
	\includegraphics[scale=0.21]{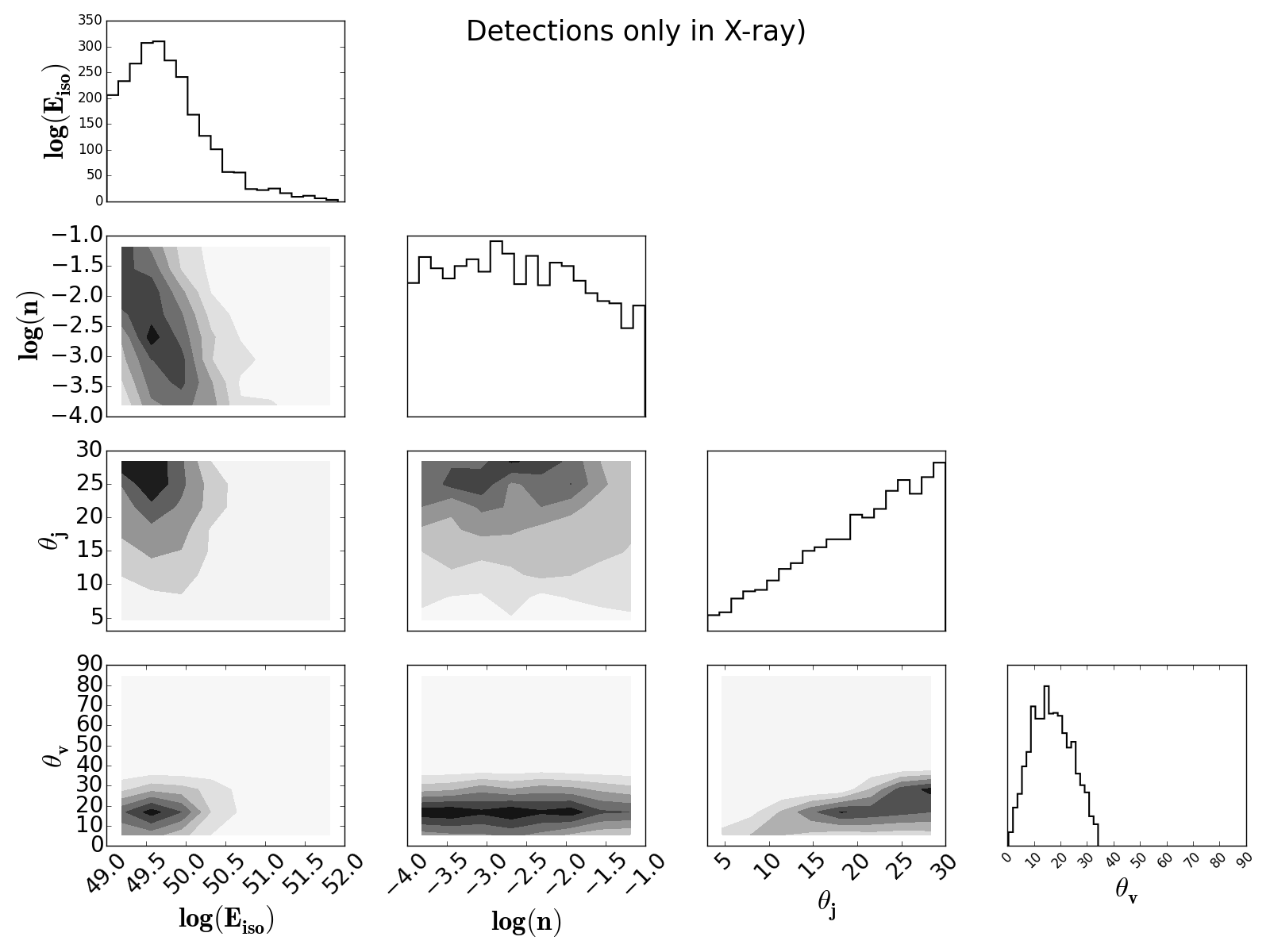}
	\includegraphics[scale=0.21]{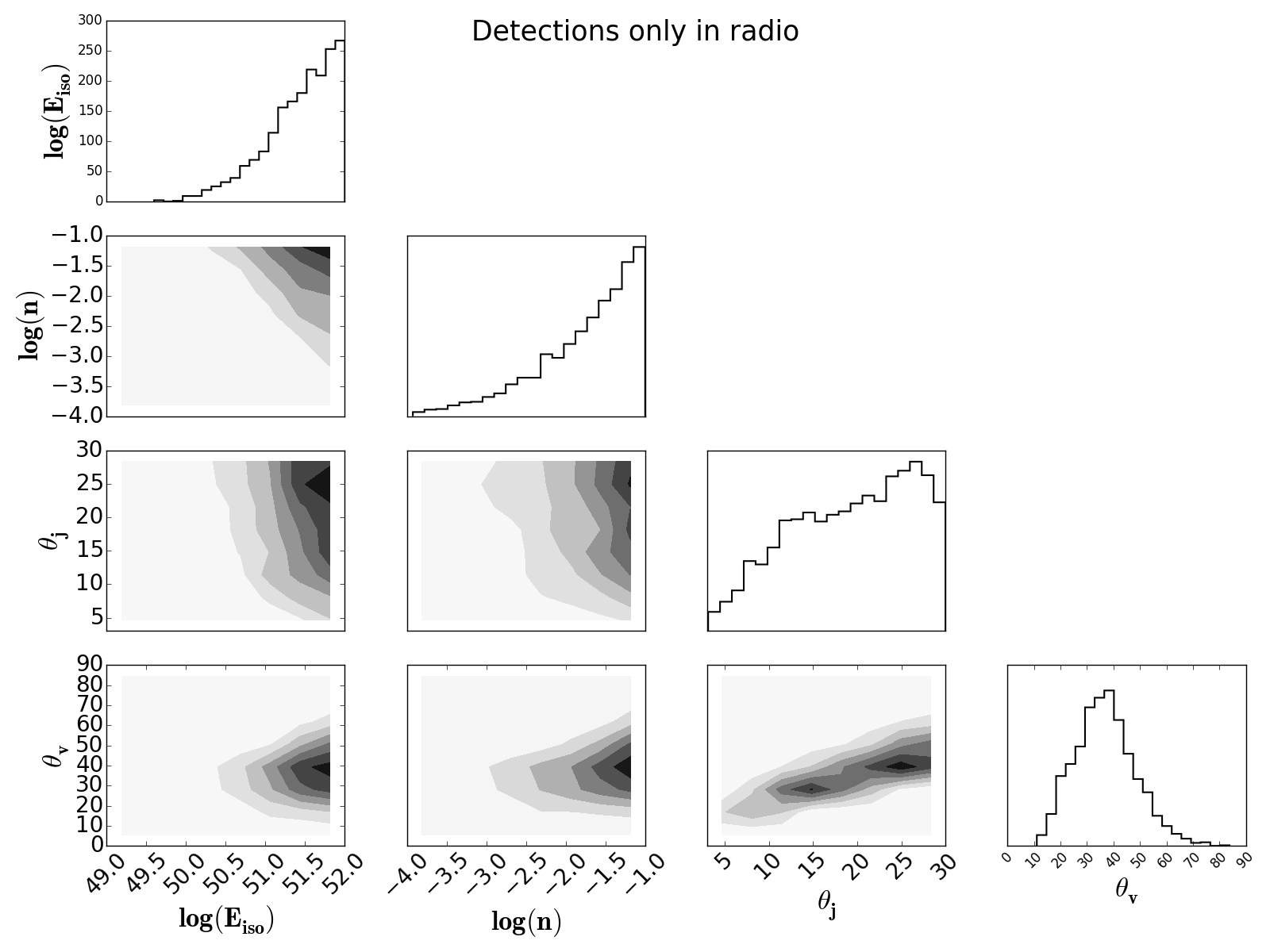}
	\includegraphics[scale=0.21]{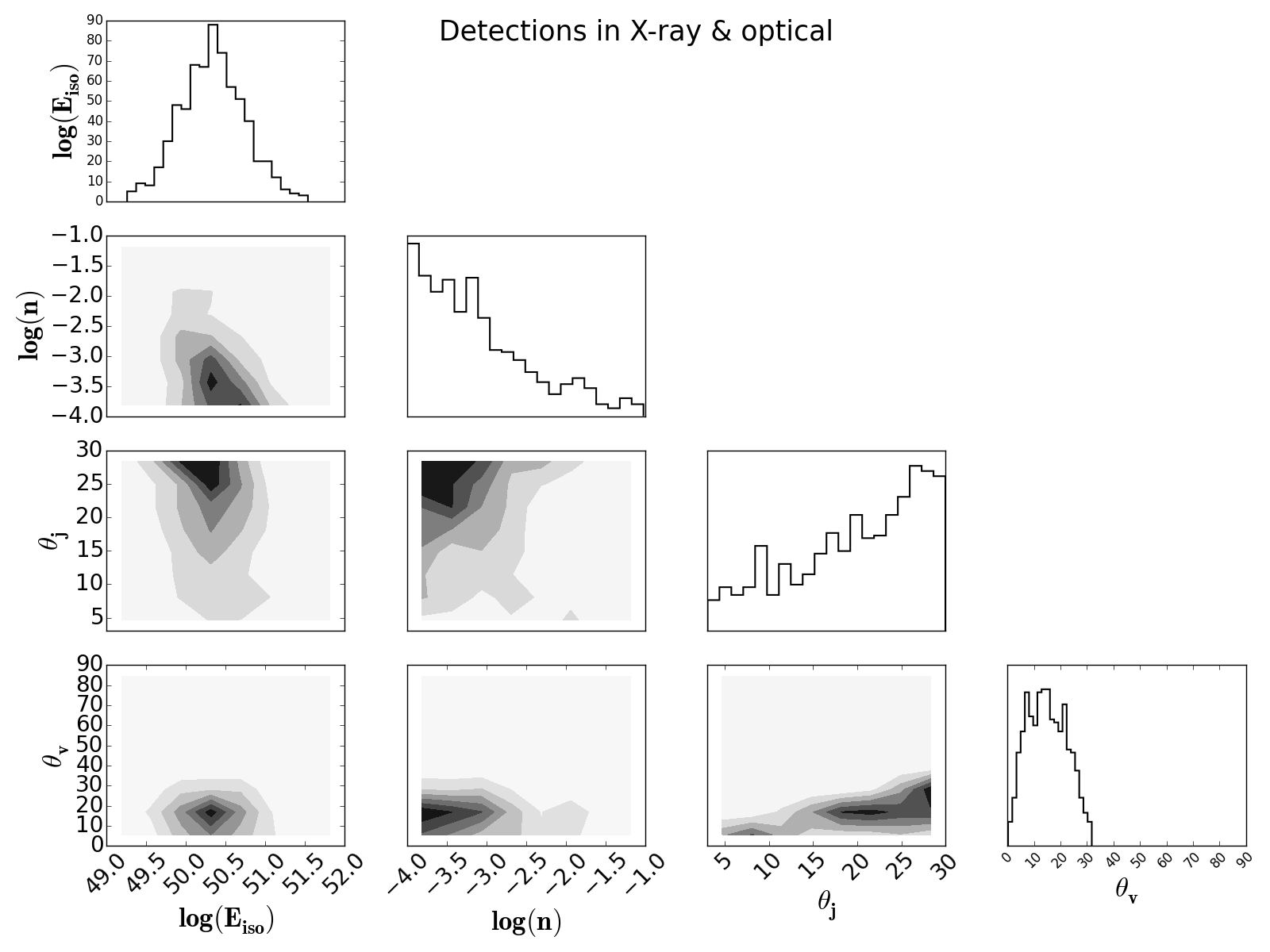}
	\includegraphics[scale=0.21]{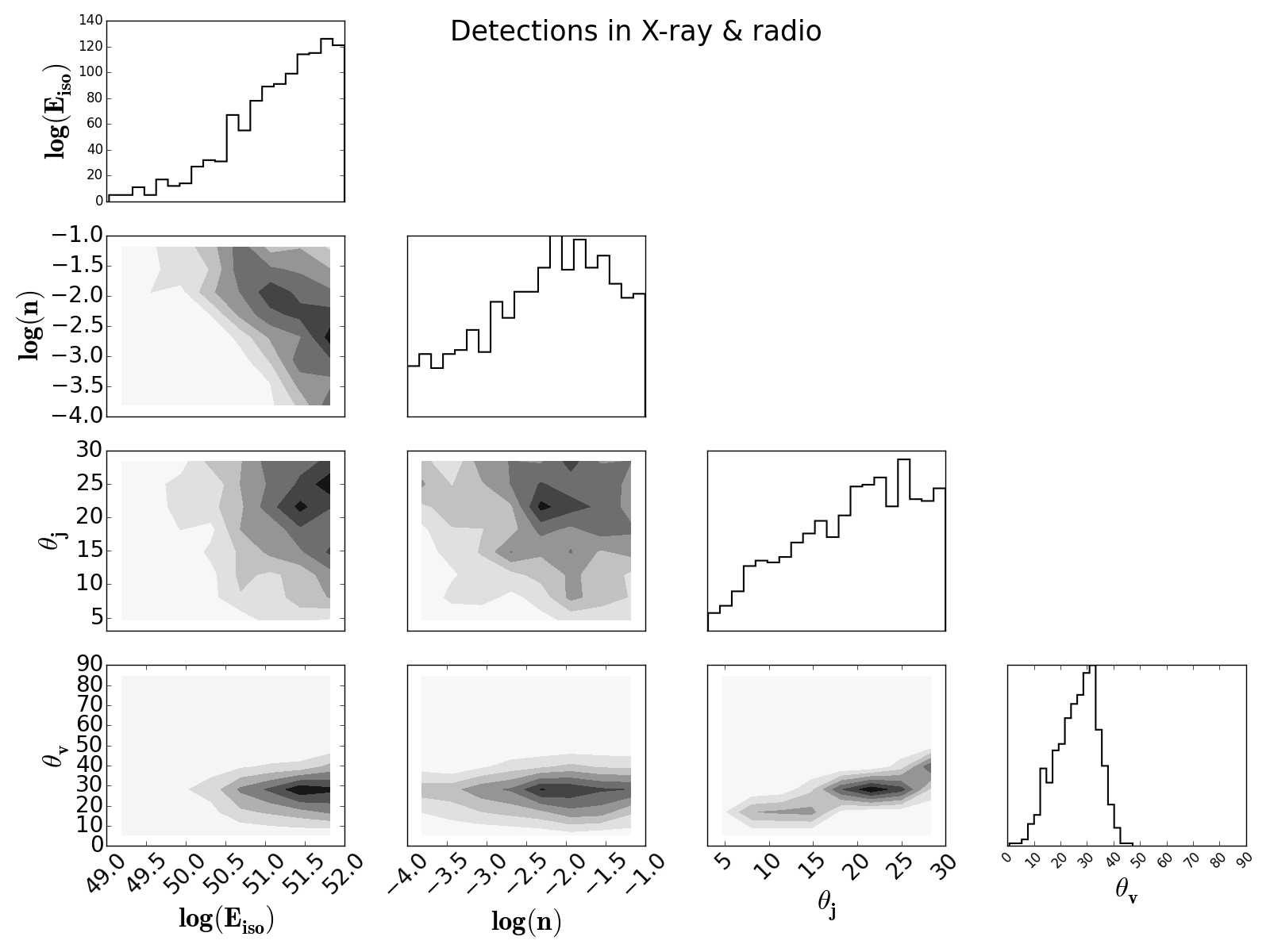}
	\caption{Combined scenarios. Upper left: Sources for which only X-ray are observed but optical and radio are not detected. Upper right: Sources for which radio is detected but X-ray and optical are not detected. Lower left: X-ray and optical are observed but radio is not detected. Lower right: X-ray and radio are observed but optical is below detection limits. In the 2D histograms, the dark region indicates higher probability.}
	\label{radiocorners}
\end{figure*}

In addition to the viewing angle effects described above, peak flux distribution is affected by the other afterglow parameters as well. Ranges of the parameters as well as their prior distributions influence the peak flux distribution. As an example, peak flux is proportional to $E_{\rm iso}$. A higher range of $E_{\rm iso}$ will lead to an increase in the $x$ and $y$ of the blue Gaussian provided the $\tj$ and $\tv$ distributions remain the same. In addition to that, if the nature of prior distribution is changed to $P(E_{\rm iso}) \propto U$ instead of $P(log(E_{\rm iso})) \propto U$, fraction of sources with higher energy will increase leading to an increase in $x$ as well as $y$. The change in distribution obviously will also affect the width of the Gaussians. This feature can be observed in the peak flux distribution of population-2 (see figure-\ref{fig-Fpeak-distr2}) which has a large fraction of high energy sources as compared to population-1.  

%{\red AN ALTERNATIVE INDEPENDENT EXPLANATION}{\blue First of all, the large area under the green curves compared to the blue ones is purely due to selection effects in our source populations. For the given GW-detectable $\tv$ distribution (figure-\ref{BNS-pop}), a uniform $\tj$ distribution drawn from the range $3-30$ degrees will naturally make a choice of outside-jet scenario ($\tv>\tj$) for $\sim85\%$ of the sources, with only the remaining $\sim15\%$ being under within-jet scenario. This reflects in the area of respective bell-curves in all 3 bands.}

%\tc{Need to check whether to include this. Or in other words, the  relatively isotropic radio emission will be detectable only if the E,n are sufficiently high(we agree with Figure.6 in \cite{metzger2012most} at this point).}

%Thus to summarise Figure.\ref{fig-Fpeak-distr}, all(or almost all) the on-axis sources are likely to be detected in both X-ray and optical bands, regardless of their energy, medium density or the distance(within the GW-detected prior ranges).
Next, we move on to analyzing the afterglow physical parameter space that favours detections in various bands.
\subsection{Detections of X-ray, optical and radio afterglows and favourable afterglow parameter space}
\label{sec42}
If SGRB observations so far have to be considered as a reference, the
detection probabilities are not the same in all bands. Our aim is to
understand the role of the afterglow physical parameters in detection
(or non-detection) in the three different bands. For this, we use
population-1 because results from population-1 appear more in
agreement with short GRB afterglow observations than population-2.

First we consider all afterglows detected in each band and analyse the
parameter distributions favouring each of them as shown in the corner
plots in Figure. \ref{fig-corner-XOR-all}. For clarity, we are
focusing only on the effect of $E_{\rm iso}, n, \tj,$ and $\tv$, and
not displaying $\eB$ which was varied in a narrow range of $0.01-0.1$. However, it has to be noted that for several afterglows $\eB$ is found to be of a lower value \citep{2009ApJ...706L..33G}. A smaller $\eB$ corresponds to a lower magnetic field of the emission region thereby the flux in all bands will be reduced. The parameters $\eE$ and $p$ were kept fixed. 

Let us first focus on the diagonal entries from all 3 corner plots of Figure. \ref{fig-corner-XOR-all} which are the 1-D histograms of the detected afterglows. It will be useful to remember that the
corresponding prior distributions of $E_{\rm iso}, n$, and $\tj$ will be a horizontal line for population-1 for which these figures are made. 
See Figure-\ref{BNS-pop} for the prior distribution of $\tv$. 
As expected, low energy afterglows fail to cross the detection threshold (upper panel
in left most column). Ambient density is an important factor for radio
afterglows while is not very significant for X-rays. This is because,
in the synchrotron spectral regime $\nu > \nu_c$ which most X-ray
afterglows are likely to occupy, $f_{\nu}$ is insensitive to $n$.
The higher probability of larger $\tj$ values is again a selection effect in our population, \textit{i.e,} for a given $\tv$ distribution, sources with larger values of $\tj$ are more likely to be within-jet and hence are more likely to be detected compared to sources with smaller values of $\tj$.
Distribution of $\tv$ extends to relatively larger values for
radio detections as expected, because the radio peaks are delayed and doppler
de-boosting is alleviated even for extreme off-axis cases by the time of their peak. (see
description in section-\ref{rad-vs-higher}).
\begin{figure*}
	\label{fig-tvj-hist-all}
	\includegraphics[scale=0.42]{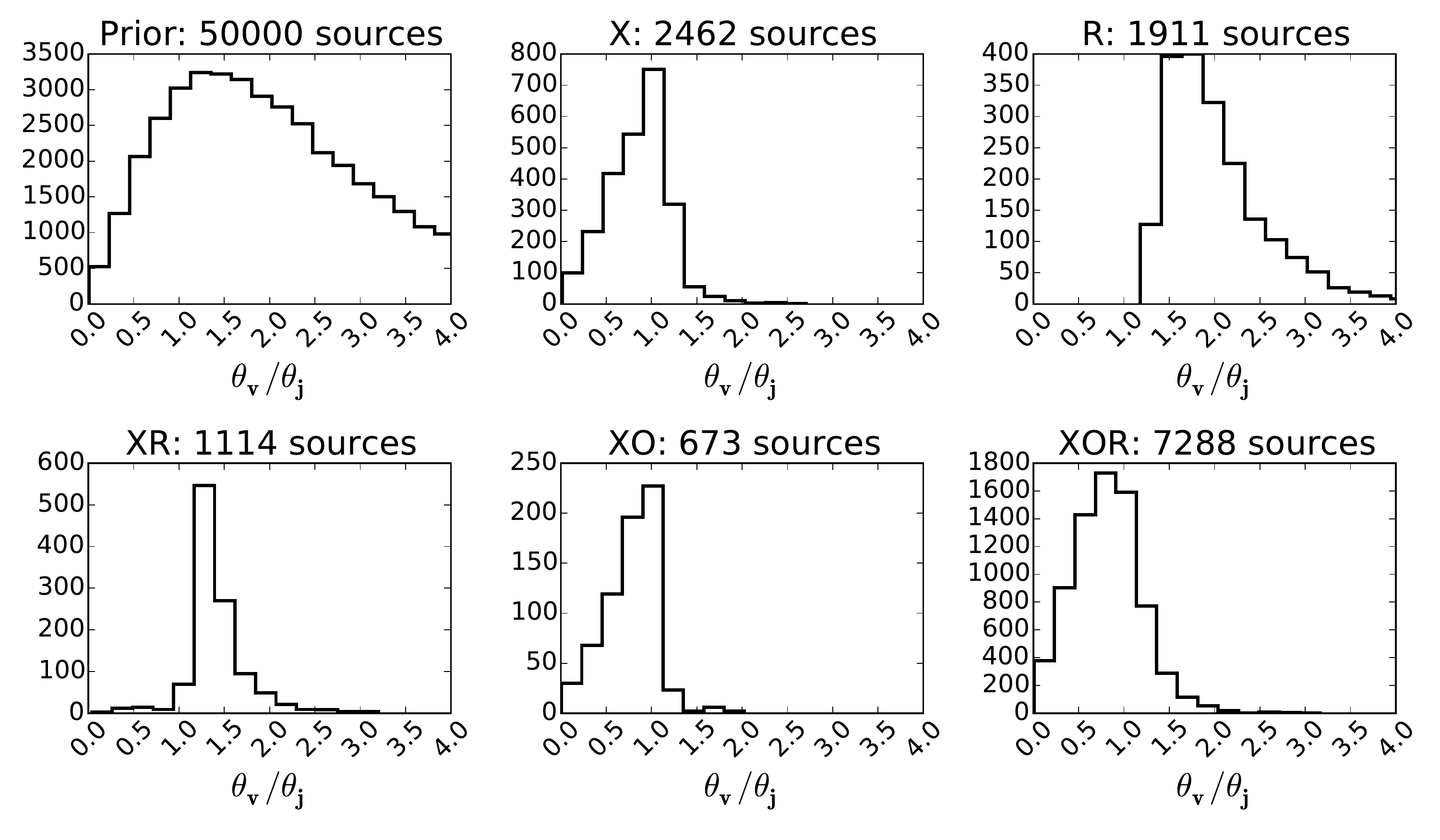}
	\caption{Distributions of ratio $\theta_v/\theta_j$ in various scenarios. The distributions show the probability for a given observer to encounter a certain scenario. For example, XO (i.e, radio alone is not detected) detection indicates that the observer is very likely to be within the jet cone. If GW observations give an estimates of $\tv$, a prediction on $\tj$ can be made using this.}
\end{figure*}

Next we explore different combinations of multi-band afterglow detection.
\subsection{Parameter space constrained for different detection scenarios}
Constraints on the afterglow parameter space can be drawn from the detections and non-detections in various bands. Here we ignore the role of cadence and field of view for non-detection in a certain band and explore the possibility that the non-detection is a consequence of afterglow parameter space and the sensitivity of instruments. For example, X-ray afterglows are detected while optical and radio afterglows are not detected for several short GRBs \citep{fong2015decade}. 

The possible combinations are: 
	\bn
		\item {X: Detection in X-ray with non-detections in  optical and radio}
		\item {O: Detection in optical with non-detections in  X-ray and radio}
		\item {R: Detection in radio with non-detections in  optical and Xray}
		\item {XO: Detections in both X-ray and optical with non-detection in radio}
		\item {XR: Detections in both X-ray and radio with non-detection in optical}
		\item {OR: Detections in both optical and radio with non-detection in X-ray}
		\item {XOR: Detections in all 3 bands - X-ray, optical and radio.}
	\en
	
We explore all combinations of observational scenarios but are presenting corner plots for a selected list which can provide insightful constraints on the afterglow parameter space. 

Further, much more stringent constraints can be arrived by using the detected flux value, which we plan to explore in future. 
%	We investigate  what are the preferred physical scenarios (ranges of distance, energy, medium density, opening angle viewing angle etc.) which can produce afterglows satisfying these observational scenarios within a given instrumental framework. Based on what have(or not) been detected, we construct probability distributions for the afterglow parameters. Note that since we do this just by considering whether a given afterglow band is observed or not, and without essentially  considering the actual properties of the observed light curves, this can not be considered as parameter estimates of the afterglow parameters. We have shown corner plots.  

\subsubsection{Detection in a single band alone}
\label{single-detection}

The top two corner plots of figure-\ref{radiocorners} shows the constrained regions in  afterglow parameter space which favour detection in a single band alone while non-detection in the remaining two bands.

The top left corner plot shows the X scenario where it is seen that $\tvj < 1$ for majority of the sources while a small fraction of outside-jet sources are also captured in this scenario (see figure-\ref{fig-tvj-hist-all}). In addition to that such a scenario arises for low energy bursts mainly, since high $E_{\rm iso}$ leads to both radio and optical detections. This can be seen by comparing the very first panels (distributions of $E_{\rm iso}$) of all 3 corner plots of Figure-\ref{fig-corner-XOR-all} where we see that for low energy bursts, X-ray detections are more likely than optical and radio. No constraint can be arrived on the number density because X-ray is not very sensitive to $n$ (see discussion in section-\ref{sec42}). 

Detection in optical alone (O scenario) is nearly difficult for the instrument thresholds we considered, \textit{ie,} the physical parameters which ensure an optical detection by \textit{LSST} always ensure an X-ray detection by \textit{XRT} also. 

A case of detection in radio band alone (R scenario) happens only if a high energy burst occurs in a high density ambient medium but viewed extremely off the jet cone ($\tvj > 1$) (see figure-\ref{radiocorners} (top right)). Because, as we have seen in figure-\ref{fig-corner-XOR-all}, high energy and high medium density are essential for radio detections. Further, being well off the jet cone ($\tvj > 1$) favours non-detection in X-ray/opt. 

\subsubsection{Detections in multiple bands}
Here, we try to identify parameter space regions which favour detections in more than one bands.

For a handful of observed short GRBs, afterglows are detected in X-ray and
optical while not detected in radio \citep{fong2015decade}. With our simulated population, we see that such a scenario (XO scenario)  is most favoured when $E_{\rm iso}$ is around $10^{50.5}$ and $n < 10^{-2}$ as shown in the bottom left panel of figure.\ref{radiocorners}. The reason is that a
non-detection in radio indicates that the energy and number density are
lower while optical detection requires the energy to be not too low. Further, to fulfil this criteria,  all these sources definitely should have $\tvj < 1$ which is essential to ensure detection in optical band as seen in figure.\ref{fig-corner-XOR-all}.
%
%Given the LSST thresholds, it is difficult to not to detect the optical afterglow. Very few sources in our simulation have optical magnitudes below the LSST threshold. {\red Resmi, IS THIS TRUE??? GIVEN THE THRESHOLDS, I THINK THIS IS ONE OF THE MOST PROBABLE SCENARIO THAT WE HAVE FOUND -- Not getting the gist, Resmi} 

An interesting scenario is when a non-detection happens in optical band alone. For this, the observer should be nearly aligned along the edge of the jet, \textit{ie,} $\tvj ~1$ (see lower right corner plot of figure-\ref{radiocorners}). This ensures that both X-ray and radio are detected and optical is not detected. Or in other words, if $\tvj <1$,  optical is likely to be detected making it XOR and if $\tvj > 1$, the X-ray flux is likely to be below XRT threshold making it an R alone case.

The most unlikely scenario is to not to detect X-ray alone (OR). Only a handful of sources in our population have satisfied this criteria. Therefore, it is impossible to obtain any meaningful constrains on the parameters from our current simulations for this condition. As discussed in section.\ref{single-detection}, this is due to the sensitivities of X-ray and optical instruments we considered. It requires a much more sensitive optical instrument than the one we considered now to enable this to be a likely scenario.

However, for a large number of within-jet cases, we get detection in all wavelengths. This is contrary to the existing short GRB observations where the radio detections are very poor. A major difference is the lower distances that our simulation is considering for GW-detected NS mergers, allowing radio fluxes to be within the VLA threshold.

\subsection{Summary of the corner plots}
The corner plots contain information of the multi-dimensional
afterglow parameter space which was concealed in the peak flux
histograms Figure-\ref{fig-Fpeak-distr}. Corner plots of XO and XOR
scenarios are populated by within-jet cases. This can be seen also in
the histograms where majority of within-jet cases are above detection
limit in both X-ray and optical bands (see the blue bell-curve in
Figure-\ref{fig-Fpeak-distr}). Therefore, for within-jet cases, for the standard ranges of afterglow physical parameters afterglow detection by XRT and LSST (or any such deep reaching optical telescopes) is ensured. Viewing from off the jet cone ($\tvj > 1$) is required to have either X-ray or optical flux to be below the corresponding sensitivity limits we have considered. However, it is difficult to have the X-ray lightcurve below the XRT threshold, if physical parameters
are ensuring an optical detection even for outside-jet cases. Hence we
do not have any optical-alone cases in our simulations. For
outside-jet, in most cases, radio afterglow will get detected. The
physical parameter space is complimentary between the R-alone and XO
cases in Figure-\ref{radiocorners}. While XO is frequent in short GRB
observations, R-alone has to wait for an NS merger triggered GRB where
the jet is viewed far off from its axis.

As a summary focusing on effects due to the viewing angle, we are presenting the different scenarios as a function of $\tvj$ in Figure-\ref{fig-tvj-hist-all}.

%%Figure.\ref{fig-tvj-hist-all} shows the Distributions of geometric profile factor $\tvj$ in all scenarios. The distributions shows what kind of observations(off-axis or on axis) are most likely if we have encountered a given scenario(X, R, XOR etc). For example, If we have XO scenario(radio missed), then we are very likely to be within the jet cone. If GW gives estimates of $\tv$, we can compute(distribution of) $\tj$ using this.  

%====================================================================

\section{Conclusions and future directions}
%(REMOVE THE ITEMIZE, ADD PARAGRAPHS: PARA1: PARADIGM, PARA2:CONTEXT, PARA3: BINARY POPULATION: pARA 4: VARIOUS SCENARIOS STUDIED, PARA 4: INSTRUMENTS ETC AND WHAT WE FOUND WITH NUMBERS ETC, PARA 6: IMPLICATIONS--RATES SHOULD BE ONE OF THE IMPLICATION AND GIVE REFERENCE TO THE LIGO DCC NO, POSSIBLE CAVEATS AND FUTURE DIRECTION COULD BE THE LAST PARAGRAPH IN WHICH YOU CAN MENTION ABOUT CADENCE ETC)

{In this work, we have explored the afterglow parameter space for coincident observations of gravitational waves and SGRB afterglows generated from BNS mergers. The detection of $\gamma$-ray emission from GRB170817A in association with gravitational waves from GW170817 along with several other EM counterparts in longer wavelengths has firmly established the association between BNS mergers and SGRBs. For BNS mergers observed in GW window, afterglows are potential EM counterparts to be followed up. In our study, we have explored the detectability and detection scenarios of afterglows in various bands irrespective of the detection of SGRB in prompt $\gamma$-ray emission.}

We simulate 50,000 GRB afterglow lightcurves, assuming them to be the Electromagnetic (EM) counterparts of Neutron Star mergers. We use the numerical hydrodynamic code \textit{BoxFit} to systematically explore the multi-dimensional afterglow parameter space. Using flux limits of three instruments, Swift-XRT, LSST, and JVLA, operating in three different bands of the EM spectrum, we explore how the afterglow parameter space results in different observational scenarios. We use the distribution of the peak flux in a given afterglow lightcurve to understand the rate of detections in that particular band, which is explored further in a companion paper \citep{saleem-etal-2017-agRates}. {{Our study focuses only on the standard forward shock driven by the GRB jet. EM counterparts from other components like reverse shock, central engine powered forward shock, merger ejecta etc. are not considered.}}

We divide the afterglow population based on the ratio ($\tvj = \tv/\tj$) of the observer's viewing angle to the jet opening angle. Within-jet sources are the ones where the observer's line of sight is within the jet cone ($\tvj  < 1$) and for outside-jet sources, the line of sight is beyond the edge of the jet ($\tvj > 1$). We find that the detection scenarios are sensitive to the ranges and distributions of the physical parameters.

We notice that most within-jet sources are detected by \textit{XRT} and \textit{LSST} (or similar deep imaging optical instruments). A non-detection in radio for within-jet sources implies relatively low jet energy and ambient medium density (roughly, $E_{\rm iso} < 10^{51}$~ergs and $n < 0.01$~atom/cc). X-ray and optical afterglows are not likely to be detected if $\tvj >>1$. However, if the jet energy and the ambient density are high enough radio afterglow alone could be detected.

In arriving at these conclusions, we have ignored the effects of field of view and cadence. In addition, the constraints on the physical parameter space are sensitive to the instruments used and detection thresholds considered. 

Here we have only considered a detection or a non-detection, but not used the detected flux value. More detailed multi-messenger astronomy can be attempted by including flux measurements.

	{Please note that the recent joint BNS merger and associated SGRB event (GW170817 and GRB170817A) was observed at a distance around 40 Mpc while the studies in this paper have considered BNS sources uniformly distributed in comoving volume above 100 Mpc. Our choice of 100 Mpc as the lower distance limit was well consistent with the existing SGRB observations until the the discovery of  GRB170817A. However, in the context of detection of GW170817+GRB170817A, given the chances of detecting nearby joint events,  we have ensured that the results and interpretations of our studies in this paper are not sensitive to this choice. This primaraly is because the binary sources are uniformly distributed in comoving volume, among the GW detections by LHVIK, only $\sim 1.4\%$ sources comes from within the 100 Mpc sphere. We have explicitly tested how the EM detectability (detection fractions of X-ray, optical and radio afterglows) changes if we use a lower distance cut-off at 20 Mpc instead of 100 Mpc.  In all the bands, the detection fractions changes by only less than $1\%$ and hence our interpretations about the detection scenarios and the associated  constraints on afterglow parameter space will remain unaffected.}

	\section{Acknowledgements}
	MS thanks to the Max Planck Partner Group HPC facilty at IISER-TVM
	and HPC facility at IUCAA, Pune where most of the numerical exercises are carried out. Development of the Boxfit code was supported in part by NASA through grant NNX10AF62G issued through the Astrophysics Theory Program and by the NSF through grant AST-1009863. Simulations for BOXFIT version 2 have been carried out in part on the computing facilities of the Computational Center for Particle and Astrophysics (C2PAP) of the research cooperation "Excellence Cluster Universe" in Garching, Germany. KGA is partially  supported by a grant
	from Infosys Foundation. AP acknowledges the IIT Bombay seed grant for the support. This document has LIGO document
        number {\tt P1700247.}
	
%====================================================================
\bibliographystyle{mnras}
\bibliography{draft}

\begin{thebibliography}{}
\makeatletter
\relax
\def\mn@urlcharsother{\let\do\@makeother \do\$\do\&\do\#\do\^\do\_\do\%\do\~}
\def\mn@doi{\begingroup\mn@urlcharsother \@ifnextchar [ {\mn@doi@}
  {\mn@doi@[]}}
\def\mn@doi@[#1]#2{\def\@tempa{#1}\ifx\@tempa\@empty \href
  {http://dx.doi.org/#2} {doi:#2}\else \href {http://dx.doi.org/#2} {#1}\fi
  \endgroup}
\def\mn@eprint#1#2{\mn@eprint@#1:#2::\@nil}
\def\mn@eprint@arXiv#1{\href {http://arxiv.org/abs/#1} {{\tt arXiv:#1}}}
\def\mn@eprint@dblp#1{\href {http://dblp.uni-trier.de/rec/bibtex/#1.xml}
  {dblp:#1}}
\def\mn@eprint@#1:#2:#3:#4\@nil{\def\@tempa {#1}\def\@tempb {#2}\def\@tempc
  {#3}\ifx \@tempc \@empty \let \@tempc \@tempb \let \@tempb \@tempa \fi \ifx
  \@tempb \@empty \def\@tempb {arXiv}\fi \@ifundefined
  {mn@eprint@\@tempb}{\@tempb:\@tempc}{\expandafter \expandafter \csname
  mn@eprint@\@tempb\endcsname \expandafter{\@tempc}}}

\bibitem[\protect\citeauthoryear{Abbott et~al.,}{Abbott
  et~al.}{2016a}]{gw150914}
Abbott B.~P.,  et~al., 2016a, Physical review letters, 116, 061102

\bibitem[\protect\citeauthoryear{Abbott et~al.,}{Abbott
  et~al.}{2016b}]{gw151226}
Abbott B.,  et~al., 2016b, Physical Review Letters, 116, 241103

\bibitem[\protect\citeauthoryear{Abbott et~al.,}{Abbott
  et~al.}{2017a}]{gw170104}
Abbott B.~P.,  et~al., 2017a, arXiv preprint arXiv:1706.01812

\bibitem[\protect\citeauthoryear{Abbott et~al.,}{Abbott
  et~al.}{2017b}]{gw170608}
Abbott B.,  et~al., 2017b, arXiv preprint arXiv:1711.05578

\bibitem[\protect\citeauthoryear{Abbott et~al.}{Abbott
  et~al.}{2017c}]{gw170814}
Abbott B.~P.,  et~al., 2017c, Submitted to: Phys. Rev. Lett.

\bibitem[\protect\citeauthoryear{Abbott et~al.,}{Abbott
  et~al.}{2017d}]{GW170817}
Abbott B.~P.,  et~al., 2017d, Physical Review Letters, 119, 161101

\bibitem[\protect\citeauthoryear{Abbott et~al.,}{Abbott
  et~al.}{2017e}]{MMApaper}
Abbott B.,  et~al., 2017e, The Astrophysical Journal Letters, 848, L13

\bibitem[\protect\citeauthoryear{Abbott et~al.,}{Abbott
  et~al.}{2017f}]{GRB+GW-2017}
Abbott B.,  et~al., 2017f, The Astrophysical Journal Letters, 848, L13

\bibitem[\protect\citeauthoryear{Akerlof et~al.}{Akerlof
  et~al.}{1999}]{Akerlof:1999aa}
Akerlof C.,  et~al., 1999, \mn@doi [Nature] {10.1038/18837}, 398, 400

\bibitem[\protect\citeauthoryear{Alexander et~al.,}{Alexander
  et~al.}{2017}]{alexander2017electromagnetic}
Alexander K.,  et~al., 2017, The Astrophysical Journal Letters, 848, L21

\bibitem[\protect\citeauthoryear{Arcavi et~al.}{Arcavi
  et~al.}{2017}]{Arcavi:2017xiz}
Arcavi I.,  et~al., 2017, \mn@doi [Nature] {10.1038/nature24291}, 551, 64

\bibitem[\protect\citeauthoryear{Arun, Tagoshi, Pai  \& Mishra}{Arun
  et~al.}{2014}]{arun2014synergy}
Arun K.,  Tagoshi H.,  Pai A.,   Mishra C.~K.,  2014, Physical Review D, 90,
  024060

\bibitem[\protect\citeauthoryear{Bartos, Brady  \& Marka}{Bartos
  et~al.}{2013}]{bartos2013gravitational}
Bartos I.,  Brady P.,   Marka S.,  2013, Classical and Quantum Gravity, 30,
  123001

\bibitem[\protect\citeauthoryear{Beniamini \& van~der Horst}{Beniamini \&
  van~der Horst}{2017}]{Beniamini:2017jil}
Beniamini P.,  van~der Horst A.~J.,  2017, arxiv

\bibitem[\protect\citeauthoryear{Berger, Fong  \& Chornock}{Berger
  et~al.}{2013}]{Berger:2013wna}
Berger E.,  Fong W.,   Chornock R.,  2013, \mn@doi [Astrophys. J.]
  {10.1088/2041-8205/774/2/L23}, 774, L23

\bibitem[\protect\citeauthoryear{Blanchet}{Blanchet}{2006}]{Bliving}
Blanchet L.,  2006, Living Rev. Rel., 9, 4

\bibitem[\protect\citeauthoryear{Blanchet, Damour, Iyer, Will  \&
  Wiseman}{Blanchet et~al.}{1995}]{BDIWW95}
Blanchet L.,  Damour T.,  Iyer B.~R.,  Will C.~M.,   Wiseman A.~G.,  1995,
  Phys. Rev. Lett., 74, 3515

\bibitem[\protect\citeauthoryear{Blanchet, Damour, Esposito-Far{\`e}se  \&
  Iyer}{Blanchet et~al.}{2004}]{BDEI04}
Blanchet L.,  Damour T.,  Esposito-Far{\`e}se G.,   Iyer B.~R.,  2004, Phys.
  Rev. Lett., 93, 091101

\bibitem[\protect\citeauthoryear{Blandford \& McKee}{Blandford \&
  McKee}{1976}]{Blandford:1976uq}
Blandford R.~D.,  McKee C.~F.,  1976, \mn@doi [Phys. Fluids]
  {10.1063/1.861619}, 19, 1130

\bibitem[\protect\citeauthoryear{Campana et~al.}{Campana
  et~al.}{2006}]{Campana:2006ja}
Campana S.,  et~al., 2006, \mn@doi [Astron. Astrophys.]
  {10.1051/0004-6361:20064856}, 454, 113

\bibitem[\protect\citeauthoryear{Costa et~al.}{Costa
  et~al.}{1997}]{Cota:1997cg}
Costa E.,  et~al., 1997, \mn@doi [Nature] {10.1038/42885}, 387, 783

\bibitem[\protect\citeauthoryear{{Dalal}, {Griest}  \& {Pruet}}{{Dalal}
  et~al.}{2002}]{2002ApJ...564..209D}
{Dalal} N.,  {Griest} K.,   {Pruet} J.,  2002, \mn@doi [\apj] {10.1086/324142},
  \href {http://adsabs.harvard.edu/abs/2002ApJ...564..209D} {564, 209}

\bibitem[\protect\citeauthoryear{Dietrich \& Ujevic}{Dietrich \&
  Ujevic}{2017}]{dietrich2017modeling}
Dietrich T.,  Ujevic M.,  2017, Classical and Quantum Gravity, 34, 105014

\bibitem[\protect\citeauthoryear{Fong, Berger  \& Fox}{Fong
  et~al.}{2010}]{Fong:2009bd}
Fong W.-f.,  Berger E.,   Fox D.~B.,  2010, \mn@doi [Astrophys. J.]
  {10.1088/0004-637X/708/1/9}, 708, 9

\bibitem[\protect\citeauthoryear{Fong et~al.}{Fong et~al.}{2013}]{Fong:2013eqa}
Fong W.-f.,  et~al., 2013, \mn@doi [Astrophys. J.]
  {10.1088/0004-637X/769/1/56}, 769, 56

\bibitem[\protect\citeauthoryear{Fong, Berger, Margutti  \& Zauderer}{Fong
  et~al.}{2015}]{fong2015decade}
Fong W.-f.,  Berger E.,  Margutti R.,   Zauderer B.~A.,  2015, The
  Astrophysical Journal, 815, 102

\bibitem[\protect\citeauthoryear{Fong et~al.}{Fong et~al.}{2016}]{Fong:2016irn}
Fong W.-f.,  et~al., 2016, \mn@doi [Astrophys. J.]
  {10.3847/1538-4357/833/2/151}, 833, 151

\bibitem[\protect\citeauthoryear{Foucart}{Foucart}{2012}]{foucart2012black}
Foucart F.,  2012, Physical Review D, 86, 124007

\bibitem[\protect\citeauthoryear{Frail, Kulkarni, Nicastro, Feroci  \&
  Taylor}{Frail et~al.}{1997}]{Frail:1997qf}
Frail D.~A.,  Kulkarni S.~R.,  Nicastro S.~R.,  Feroci M.,   Taylor G.~B.,
  1997, \mn@doi [Nature] {10.1038/38451}, 389, 261

\bibitem[\protect\citeauthoryear{Frontera et~al.}{Frontera
  et~al.}{1998}]{Frontera:1997ae}
Frontera F.,  et~al., 1998, \mn@doi [Astrophys. J.] {10.1086/311132}, 493, L67

\bibitem[\protect\citeauthoryear{{Gao}, {Mao}, {Xu}  \& {Fan}}{{Gao}
  et~al.}{2009}]{2009ApJ...706L..33G}
{Gao} W.-H.,  {Mao} J.,  {Xu} D.,   {Fan} Y.-Z.,  2009, \mn@doi [\apjl]
  {10.1088/0004-637X/706/1/L33}, \href
  {http://adsabs.harvard.edu/abs/2009ApJ...706L..33G} {706, L33}

\bibitem[\protect\citeauthoryear{{Gehrels}, {Ramirez-Ruiz}  \& {Fox}}{{Gehrels}
  et~al.}{2009}]{2009ARA&A..47..567G}
{Gehrels} N.,  {Ramirez-Ruiz} E.,   {Fox} D.~B.,  2009, \mn@doi [\araa]
  {10.1146/annurev.astro.46.060407.145147}, \href
  {http://adsabs.harvard.edu/abs/2009ARA%26A..47..567G} {47, 567}

\bibitem[\protect\citeauthoryear{Giacomazzo, Perna, Rezzolla, Troja  \&
  Lazzati}{Giacomazzo et~al.}{2012}]{giacomazzo2012compact}
Giacomazzo B.,  Perna R.,  Rezzolla L.,  Troja E.,   Lazzati D.,  2012, The
  Astrophysical Journal Letters, 762, L18

\bibitem[\protect\citeauthoryear{Goodman}{Goodman}{1986}]{Goodman:1986az}
Goodman J.,  1986, \mn@doi [Astrophys. J.] {10.1086/184741}, 308, L47

\bibitem[\protect\citeauthoryear{Granot, Panaitescu, Kumar  \& Woosley}{Granot
  et~al.}{2002}]{Granot:2002za}
Granot J.,  Panaitescu A.,  Kumar P.,   Woosley S.~E.,  2002, \mn@doi
  [Astrophys. J.] {10.1086/340991}, 570, L61

\bibitem[\protect\citeauthoryear{Grupe, Burrows, Patel, Kouveliotou, Zhang,
  Meszaros, Wijers  \& Gehrels}{Grupe et~al.}{2006}]{Grupe:2006uc}
Grupe D.,  Burrows D.~N.,  Patel S.~K.,  Kouveliotou C.,  Zhang B.,  Meszaros
  P.,  Wijers R. A.~M.,   Gehrels N.,  2006, \mn@doi [Astrophys. J.]
  {10.1086/508739}, 653, 462

\bibitem[\protect\citeauthoryear{Guetta \& Piran}{Guetta \&
  Piran}{2005}]{Guetta:2004fc}
Guetta D.,  Piran T.,  2005, \mn@doi [Astron. Astrophys.]
  {10.1051/0004-6361:20041702}, 435, 421

\bibitem[\protect\citeauthoryear{Hallinan et~al.,}{Hallinan
  et~al.}{2017}]{hallinan2017radio}
Hallinan G.,  et~al., 2017, Science, p. eaap9855

\bibitem[\protect\citeauthoryear{Harrison et~al.}{Harrison
  et~al.}{1999}]{Harrison:1999hv}
Harrison F.~A.,  et~al., 1999, \mn@doi [Astrophys. J.] {10.1086/312282}, 523,
  L121

\bibitem[\protect\citeauthoryear{Hjorth et~al.}{Hjorth
  et~al.}{2003}]{Hjorth:2003jt}
Hjorth J.,  et~al., 2003, \mn@doi [Nature] {10.1038/nature01750}, 423, 847

\bibitem[\protect\citeauthoryear{Kawaguchi, Kyutoku, Shibata  \&
  Tanaka}{Kawaguchi et~al.}{2016}]{kawaguchi2016}
Kawaguchi K.,  Kyutoku K.,  Shibata M.,   Tanaka M.,  2016, The Astrophysical
  Journal, 825, 52

\bibitem[\protect\citeauthoryear{Kumar \& Zhang}{Kumar \&
  Zhang}{2014}]{Kumar:2014upa}
Kumar P.,  Zhang B.,  2014, \mn@doi [Phys. Rept.]
  {10.1016/j.physrep.2014.09.008}, 561, 1

\bibitem[\protect\citeauthoryear{Levan et~al.}{Levan
  et~al.}{2008}]{Levan:2007zr}
Levan A.~J.,  et~al., 2008, \mn@doi [Mon. Not. Roy. Astron. Soc.]
  {10.1111/j.1365-2966.2007.11953.x}, 384, 541

\bibitem[\protect\citeauthoryear{Li \& Paczynski}{Li \&
  Paczynski}{1998}]{Li:1998bw}
Li L.-X.,  Paczynski B.,  1998, \mn@doi [Astrophys. J.] {10.1086/311680}, 507,
  L59

\bibitem[\protect\citeauthoryear{Margutti et~al.}{Margutti
  et~al.}{2017}]{Margutti:2017cjl}
Margutti R.,  et~al., 2017, \mn@doi [Astrophys. J.] {10.3847/2041-8213/aa9057},
  848, L20

\bibitem[\protect\citeauthoryear{{M{\'e}sz{\'a}ros}}{{M{\'e}sz{\'a}ros}}{2006}]{2006RPPh...69.2259M}
{M{\'e}sz{\'a}ros} P.,  2006, \mn@doi [Reports on Progress in Physics]
  {10.1088/0034-4885/69/8/R01}, \href
  {http://adsabs.harvard.edu/abs/2006RPPh...69.2259M} {69, 2259}

\bibitem[\protect\citeauthoryear{{Meszaros} \& {Rees}}{{Meszaros} \&
  {Rees}}{1993}]{1993ApJM}
{Meszaros} P.,  {Rees} M.~J.,  1993, \mn@doi [\apj] {10.1086/172360}, \href
  {http://adsabs.harvard.edu/abs/1993ApJ...405..278M} {405, 278}

\bibitem[\protect\citeauthoryear{Meszaros \& Rees}{Meszaros \&
  Rees}{1997}]{Meszaros:1996sv}
Meszaros P.,  Rees M.~J.,  1997, \mn@doi [Astrophys. J.] {10.1086/303625}, 476,
  232

\bibitem[\protect\citeauthoryear{Moderski, Sikora  \& Bulik}{Moderski
  et~al.}{2000}]{Moderski:1999ct}
Moderski R.,  Sikora M.,   Bulik T.,  2000, \mn@doi [Astrophys. J.]
  {10.1086/308257}, 529, 151

\bibitem[\protect\citeauthoryear{Paczynski}{Paczynski}{1986}]{Paczynski:1986px}
Paczynski B.,  1986, \mn@doi [Astrophys. J.] {10.1086/184740}, 308, L43

\bibitem[\protect\citeauthoryear{Paczynski \& Rhoads}{Paczynski \&
  Rhoads}{1993}]{Paczynski:1993gz}
Paczynski B.,  Rhoads J.~E.,  1993, \mn@doi [Astrophys. J.] {10.1086/187102},
  418, L5

\bibitem[\protect\citeauthoryear{Pai, Dhurandhar  \& Bose}{Pai
  et~al.}{2001}]{Pai2000}
Pai A.,  Dhurandhar S.,   Bose S.,  2001, \mn@doi [Phys. Rev.]
  {10.1103/PhysRevD.64.042004}, D64, 042004

\bibitem[\protect\citeauthoryear{Panaitescu \& Kumar}{Panaitescu \&
  Kumar}{2000}]{Panaitescu:2000bk}
Panaitescu A.,  Kumar P.,  2000, \mn@doi [Astrophys. J.] {10.1086/317090}, 543,
  66

\bibitem[\protect\citeauthoryear{Panaitescu \& Kumar}{Panaitescu \&
  Kumar}{2001a}]{Panaitescu:2000xk}
Panaitescu A.,  Kumar P.,  2001a, \mn@doi [Astrophys. J.] {10.1086/321388},
  554, 667

\bibitem[\protect\citeauthoryear{Panaitescu \& Kumar}{Panaitescu \&
  Kumar}{2001b}]{Panaitescu:2001fv}
Panaitescu A.,  Kumar P.,  2001b, \mn@doi [Astrophys. J.] {10.1086/324061},
  560, L49

\bibitem[\protect\citeauthoryear{Pian et~al.}{Pian et~al.}{2017}]{Pian:2017gtc}
Pian E.,  et~al., 2017, \mn@doi [Nature] {10.1038/nature24298}, 551, 67

\bibitem[\protect\citeauthoryear{{Piran}}{{Piran}}{1999}]{1999PhR...314..575P}
{Piran} T.,  1999, \mn@doi [\physrep] {10.1016/S0370-1573(98)00127-6}, \href
  {http://adsabs.harvard.edu/abs/1999PhR...314..575P} {314, 575}

\bibitem[\protect\citeauthoryear{{Rees} \& {Meszaros}}{{Rees} \&
  {Meszaros}}{1992}]{1992MNRAS.258P..41R}
{Rees} M.~J.,  {Meszaros} P.,  1992, \mn@doi [\mnras]
  {10.1093/mnras/258.1.41P}, \href
  {http://adsabs.harvard.edu/abs/1992MNRAS.258P..41R} {258, 41P}

\bibitem[\protect\citeauthoryear{Resmi \& Bhattacharya}{Resmi \&
  Bhattacharya}{2008}]{Resmi:2008qb}
Resmi L.,  Bhattacharya D.,  2008, \mn@doi [Mon. Not. Roy. Astron. Soc.]
  {10.1111/j.1365-2966.2008.13298.x}, 388, 144

\bibitem[\protect\citeauthoryear{Resmi et~al.}{Resmi
  et~al.}{2005}]{Resmi:2005bj}
Resmi L.,  et~al., 2005, \mn@doi [Astron. Astrophys.]
  {10.1051/0004-6361:20041642}, 440, 477

\bibitem[\protect\citeauthoryear{Rezzolla, Giacomazzo, Baiotti, Granot,
  Kouveliotou  \& Aloy}{Rezzolla et~al.}{2011}]{rezzolla2011missing}
Rezzolla L.,  Giacomazzo B.,  Baiotti L.,  Granot J.,  Kouveliotou C.,   Aloy
  M.~A.,  2011, The Astrophysical Journal Letters, 732, L6

\bibitem[\protect\citeauthoryear{Rhoads}{Rhoads}{1999}]{Rhoads:1999wm}
Rhoads J.~E.,  1999, \mn@doi [Astrophys. J.] {10.1086/307907}, 525, 737

\bibitem[\protect\citeauthoryear{Rossi, Lazzati  \& Rees}{Rossi
  et~al.}{2002}]{Rossi:2001pk}
Rossi E.,  Lazzati D.,   Rees M.~J.,  2002, \mn@doi [Mon. Not. Roy. Astron.
  Soc.] {10.1046/j.1365-8711.2002.05363.x}, 332, 945

\bibitem[\protect\citeauthoryear{Rybicki \& Lightman}{Rybicki \&
  Lightman}{1979}]{rybicki1979radiative}
Rybicki G.,  Lightman A.,  1979, Radiative Processes in Astrophysics.
A Wiley-Interscience publication, Wiley, \url
  {https://books.google.co.in/books?id=LtdEjNABMlsC}

\bibitem[\protect\citeauthoryear{Saleem, Pai, Misra, Resmi  \& Arun}{Saleem
  et~al.}{2017}]{saleem-etal-2017-agRates}
Saleem M.,  Pai A.,  Misra K.,  Resmi L.,   Arun K.,  2017, arXiv preprint
  arXiv:1710.06111 (Submitted to MNRAS)

\bibitem[\protect\citeauthoryear{{Sari}}{{Sari}}{1997}]{1997ApJ...489L..37S}
{Sari} R.,  1997, \mn@doi [\apjl] {10.1086/310957}, \href
  {http://adsabs.harvard.edu/abs/1997ApJ...489L..37S} {489, L37}

\bibitem[\protect\citeauthoryear{Sari \& Piran}{Sari \&
  Piran}{1999}]{Sari:1999kj}
Sari R.,  Piran T.,  1999, \mn@doi [Astrophys.J.] {10.1086/307508}, 520, 641

\bibitem[\protect\citeauthoryear{Sari, Narayan  \& Piran}{Sari
  et~al.}{1996}]{Sari:1996za}
Sari R.,  Narayan R.,   Piran T.,  1996, \mn@doi [Astrophys. J.]
  {10.1086/178136}, 473, 204

\bibitem[\protect\citeauthoryear{Sari, Piran  \& Narayan}{Sari
  et~al.}{1998}]{Sari:1997qe}
Sari R.,  Piran T.,   Narayan R.,  1998, \mn@doi [Astrophys. J.]
  {10.1086/311269}, 497, L17

\bibitem[\protect\citeauthoryear{Sari, Piran  \& Halpern}{Sari
  et~al.}{1999}]{Sari:1999mr}
Sari R.,  Piran T.,   Halpern J.,  1999, \mn@doi [Astrophys. J.]
  {10.1086/312109}, 519, L17

\bibitem[\protect\citeauthoryear{Smartt et~al.}{Smartt
  et~al.}{2017}]{Smartt:2017fuw}
Smartt S.~J.,  et~al., 2017, \mn@doi [Nature] {10.1038/nature24303}, 551, 75

\bibitem[\protect\citeauthoryear{Tanvir, Levan, Fruchter, Hjorth, Wiersema,
  Tunnicliffe  \& de Ugarte~Postigo}{Tanvir et~al.}{2013}]{Tanvir:2013pia}
Tanvir N.~R.,  Levan A.~J.,  Fruchter A.~S.,  Hjorth J.,  Wiersema K.,
  Tunnicliffe R.,   de Ugarte~Postigo A.,  2013, \mn@doi [Nature]
  {10.1038/nature12505}, 500, 547

\bibitem[\protect\citeauthoryear{Taylor, Frail, Berger  \& Kulkarni}{Taylor
  et~al.}{2004}]{Taylor:2004wd}
Taylor G.~B.,  Frail D.~A.,  Berger E.,   Kulkarni S.~R.,  2004, \mn@doi
  [Astrophys. J.] {10.1086/422554}, 609, L1

\bibitem[\protect\citeauthoryear{Troja et~al.,}{Troja
  et~al.}{2017}]{troja2017x}
Troja E.,  et~al., 2017, Nature, 551, nature24290

\bibitem[\protect\citeauthoryear{Waxman}{Waxman}{1997}]{Waxman:1997ga}
Waxman E.,  1997, \mn@doi [Astrophys. J.] {10.1086/310960}, 489, L33

\bibitem[\protect\citeauthoryear{Wijers \& Galama}{Wijers \&
  Galama}{1999}]{Wijers:1998st}
Wijers R. A. M.~J.,  Galama T.~J.,  1999, \mn@doi [Astrophys. J.]
  {10.1086/307705}, 523, 177

\bibitem[\protect\citeauthoryear{van Eerten, Leventis, Meliani, Wijers  \&
  Keppens}{van Eerten et~al.}{2010a}]{vanEerten:2009pa}
van Eerten H.~J.,  Leventis K.,  Meliani Z.,  Wijers R. A. M.~J.,   Keppens R.,
   2010a, \mn@doi [Mon. Not. Roy. Astron. Soc.]
  {10.1111/j.1365-2966.2009.16109.x}, 403, 300

\bibitem[\protect\citeauthoryear{van Eerten, Zhang  \& MacFadyen}{van Eerten
  et~al.}{2010b}]{vanEerten:2010zh}
van Eerten H.,  Zhang W.,   MacFadyen A.,  2010b, \mn@doi [Astrophys. J.]
  {10.1088/0004-637X/722/1/235}, 722, 235

\bibitem[\protect\citeauthoryear{van Paradijs et~al.}{van Paradijs
  et~al.}{1997}]{vanParadijs:1997wr}
van Paradijs J.,  et~al., 1997, \mn@doi [Nature] {10.1038/386686a0}, 386, 686

\makeatother
\end{thebibliography}
\end{document}